\documentstyle[aps,eqsecnum,preprint,floats,epsf]{revtex}

\begin{document}
\tighten
\draft

\title{
\Large\bf Effective Field Theory of Heavy Mesons}

\author{{\bf Hai-Yang Cheng, Chi-Yee Cheung, and Wei-Min Zhang} \\
Institute of Physics, Academia Sinica, Taipei, Taiwan 115, R.O.C.}

\date{April 15, 1998}

\maketitle

\begin{abstract}
In this paper we present a detailed formulation for a recently
proposed effective field theory to describe the nonperturbative 
QCD dynamics of heavy mesons.  This effective theory incorporates 
with heavy quark symmetry (HQS) and the heavy quark effective 
theory (HQET). Heavy mesons in this theory are constructed as 
composite particles of a heavy quark bounded with the light degrees 
of freedom. The heavy meson properties in the heavy quark limit and 
the $1/m_Q$ corrections can then be explicitly evaluated 
from this effective theory. 
All the basic parameters of the HQET, namely, the heavy quark mass $m_Q$, 
the heavy meson residual mass $\overline{\Lambda}$, and the HQS 
breaking mass parameters $\lambda_1$ and $\lambda_2$, are consistently 
determined.  $\lambda_1$ is found to be small due to a 
large cancellation between the heavy quark kinetic energy and the 
chromo-electric interaction between the heavy quark and light 
degrees of freedom. We also evaluate the Isgur-Wise function, the decay 
constant, and the axial-vector coupling constant of heavy mesons. 
\end{abstract}

\vspace{0.5in}

\pacs{PACS numbers: 12.38.-t, 12.39.Hg, 12.39.Ki, 12.60.Rc, 14.40.-n}

\newpage

\section{Introduction}
 
Within the last decade, the most significant progress in the 
QCD description of hadronic physics is the discovery of heavy 
quark symmetry (HQS) \cite{IW89}, which, together with the 
development of heavy quark effective theory (HQET) from QCD 
\cite{Georgi90}, has largely simplified the analysis of heavy 
hadron physics \cite{Neubert94}. However, even in the infinite 
quark mass limit, the general properties of heavy hadrons, 
namely, their decay constants, transition form factors and 
structure functions etc. measured in various exclusive and 
inclusive decay processes remain unknown in the framework 
of QCD. The difficulty in understanding the heavy hadronic 
properties essentially lie in nonperturbative QCD dynamics, 
but HQET itself does not provide such a description to 
nonperturbative QCD. Lattice QCD simulations \cite{lattice} 
permit a nonperturbative approach to the low-energy QCD problem, 
but so far a direct lattice calculation with heavy quarks is 
still not possible due to the difficulty of placing heavy 
particles on the lattice. An alternative first-principles 
calculation of nonperturbative QCD to these heavy hadron 
properties is to solve the heavy meson bound states directly 
from light-front QCD \cite{zhang97}, but to obtain reliable results, 
further investigations are needed. Therefore, in most recent 
studies, these heavy hadron properties are usually evaluated 
using phenomenological models, such as the constituent quark 
model \cite{Isgur89}, the MIT bag model\cite{MIT}, the QCD 
sum rule \cite{Ball91}, and the light-front quark model 
\cite{Te76}.

As is well known, although the constituent quark model (CQM) 
and the MIT bag model have been widely used in the phenomenological 
discussion of hadronic structures, applications of these two models 
are trustworthy only for processes involving small momentum transfers. 
This is due mainly to the nonrelativistic limitation of the CQM and 
to the difficulty with boost in the bag model. The light-front quark 
model (LFQM) which is a relativistic quark model with simple boost 
kinematics allows us to describe physical processes with large 
momentum transfers, but it is still not truely Lorentz covariant due to the
exclusion of the so-called $Z$-diagrams \cite{dubin93,cheung97}.  As a 
result, certain theoretical ambiguities arise in LFQM calculation
which may lead to inconsistent results as have been shown in 
\cite{cheng97a,demchuk97}.  

To overcome the drawbacks in these phenomenological models, we 
recently proposed a covariant light-front model \cite{cheng97b} 
which modifies the conventional LFQM in the heavy quark limit by 
adding a constraint on the light-front wave function. The covariant 
light-front model rules out some non-covariant light-front wave 
function often used in the literature, and therefore partially 
eliminates the ambiguities presented in previous calculations. 
Meanwhile, it also largely simplifies the light-front formulation, 
and may further provide a first-principles QCD analysis of the 
$1/m_Q$ corrections within HQET \cite{cheng97b}.

However, further investigation indicates that although it has 
overcome some theoretical difficulties encountered in the conventional 
light-front formulation for heavy hadrons, our covariant quark 
model \cite{cheng97b} still cannot provide fully consistent 
results for processes involving light quark currents. This is because 
for those processes involving only heavy quark transitions, the
$Z$-diagram contributions are suppressed in the heavy quark limit;
however, when the light quarks (or currents) are involved, the 
non-covariant light-front treatment of light quarks (due to lack of 
$Z$-diagram contributions from the light quark production) will
still cause theoretical ambiguities.

In this paper, we will present a detailed formulation of a fully 
field theoretical description of heavy mesons in terms of an 
effective theory of composite particles we have proposed very 
recently\cite{cheung97a}. The composite 
particle consists of a reduced heavy quark (heavy quark 
in the heavy quark limit) coupled with the 
light degrees of freedom, in which a structure function of the composite 
particle, $\Psi(v \cdot p_q)$, corresponding to the wave function 
of a heavy meson bound state, is explicitly built in. In this field 
theory description, the constituents in composite particles are no 
longer fixed at a given instant point of time, which is a condition
imposed in  the usual construction of hadronic bound states in various 
quark models. The field theory structure of the constituents in hadrons 
allows us to formulate the physical processes in terms of the standard 
Feynman diagrams in which various time-ordering diagrams are all 
automatically included. Therefore, the lack of $Z$-diagrams in the 
usual quark 
model descriptions is no longer a problem in the present formulation. 

Moreover, combining the effective Lagrangian of the composite heavy 
mesons with the $1/m_Q$ expansion of the heavy quark QCD Lagrangian, 
we can systematically evaluate various $1/m_Q$ corrections to heavy 
meson properties in the standard framework of perturbative field 
theory. Thus, a self-contained description of heavy mesons (including 
the bound state structure and $1/m_Q$ corrections) is realized
in a field-theoretic framework.  
This effective field theory allows us to explore 
the nonperturbative heavy meson dynamics, which is not possible in the 
light-front quark models. 

The rest of the paper is organized as follows: In Sec.~II, we will
analyze the basic structure of heavy mesons in the heavy quark limit
as a composite particle of the reduced heavy quark coupled with the 
light degrees of freedom. In Sec.~III, we construct an effective 
Lagrangian to describe the composite structure of heavy mesons and 
combine this effective Lagrangian with HQET to establish a realistic 
effective field theory which can be used to evaluate the $1/m_Q$ 
corrections in the standard Feynman diagrammatic approach.  
As first applications, we evaluate the Isgur-Wise function, the heavy 
meson decay constant and the axial-vector coupling constant in the heavy 
quark limit in Sec.~IV. In  Sec.~V, we compute the pseudoscalar and 
vector heavy meson masses up to $1/m_Q$, which determine the HQET 
parameters $\lambda_1$ and $\lambda_2$. In Sec.~VI, we 
compare the effective field theory with the covariant light-front model 
\cite{cheng97b}, and show how the effective field theory overcomes the 
lack of relativistic covariance in light-front quark 
models. In Sec.~VII, we present some numerical calculations to check the
self-consistency of the theory, and then calculate all the basic 
parameters in HQET. The summary and perspective are given in Sec.~VIII.

\section{A Composite Particle Picture of Heavy Mesons in the heavy
quark limit}

We define the general expression for the composite operators of  
pseudoscalar and vector heavy mesons as follows: 
\begin{equation}
	{\bf H}_{ci}(X) = \int d^4 y \overline{q}(X-\alpha y) 
        \Gamma_i Q[X+(1-\alpha) y] ~,
	~~~ i =P, V  \, ,
\end{equation}
where the subscript $c$ means ``composite", $Q$ and $q$ are the heavy and 
light quark field operators respectively, $X$ is the center-of-mass
coordinate of the heavy meson and $y=x_Q-x_q$ the relative coordinate 
between the heavy and light quarks, $\alpha=m_Q/(m_Q+m_q)$, and 
$\Gamma_P = \gamma_5$ and $ \Gamma_V=\gamma^\mu $ 
define the spin structures for the pseudoscalar $(0^-$) meson and 
vector ($1^-$) meson, respectively. 
Inside the heavy meson, QCD dynamics is nonperturbative, so that $Q(x)$ 
and $\overline{q}(x)$ are strongly coupled, and they are 
surrounded by infinite number of $q\overline{q}$ 
pairs and gluons originating from the nontrivial QCD vacuum. 
We may phenomenologically rewrite 
the above composite operator of heavy mesons in terms of the 
constituent valence heavy and light quark field operators, denoted by $Q_0$ 
and $q_0$ respectively, coupled through a structure function $F(y)$ which 
describes the binding effect of infinite number of $q\overline{q}$ 
pairs and gluons governed by QCD,
\begin{equation}
	{\bf H}_{ci}(X) = \int d^4y \overline{q}_0(X-\alpha y) F(y) 
		\Gamma_i Q_0[X+(1-\alpha)y] \, . \end{equation}

Now we take the heavy quark limit, i.e., $m_Q \rightarrow \infty$. 
In the momentum space, 
the heavy meson carries momentum $P^\mu \equiv m_Qv^\mu + 
p_H^\mu$, where $v^\mu$ is a four velocity $(v^2=1)$ of the heavy 
meson, and $p_H$ the residual momentum. The heavy meson
on-mass-shell condition $P^\mu_H = M_H v^\mu$ corresponds to
$p_H^\mu= \overline{\Lambda} v^\mu$, where $M_H$ is the heavy 
meson mass which approaches to infinity under the heavy 
quark limit but $\overline{\Lambda} = M_H - m_Q$ is kept 
finite. Note that $\overline{\Lambda}$ is a basic parameter 
in HQET known as the residual mass of the heavy meson.
 
In the heavy quark limit, we define the so-called reduced 
heavy quark field $h_v$ in the heavy quark expansion\cite{Georgi90}, 
\begin{equation} \label{hqe}
	Q_0(x) = e^{-im_Q v \cdot x} h_v + O(1/m_Q) ~~, ~~~
		{1 + \not \! v \over 2} h_v = h_v  \, .
\end{equation}
We also introduce the reduced heavy meson composite operator $H_{ci}$,
\begin{equation}  \label{rshm}
	{\bf H}_{ci}(X) = {1\over \sqrt{M_H}} e^{-im_Qv \cdot X}
		H_{ci}(X)  \, .
\end{equation} 
Then the Fourier transformation of the composite field operator in the 
momentum space can be expressed as 
\begin{eqnarray}
	H_{ci}(v,p_H) &=& \int d^4 X e^{ip_H \cdot X} H_{ci}(X) 
			\nonumber \\
	       &=& \int {d^4 k \over (2\pi)^4} {d^4 p_q \over
		   (2\pi)^4} (2\pi)^4 \delta^4 (p_H - k - p_q) \nonumber \\
	     & & ~~~~~~~~~~~~~~\times \overline{q}_0(p_q) \Psi(v \cdot p_q)
		   \Gamma_i h_v(k) \, , \label{mhmf}
\end{eqnarray}
where $k^\mu=p_Q^\mu - m_Q v^\mu$ is the residual momentum of the heavy 
quark, and $p_q$ the momentum of the light antiquark. The function $\Psi 
(v \cdot p_q)$ is the heavy-quark-limit expression of the Fourier 
transformation of $F(y)$, which is the analog of heavy meson wave 
function in the covariant light-front quark model \cite{cheng97b}.

One may ask why is $\Psi ( v\cdot p_q)$ only a function of $v \cdot
p_q$?  Note that the Fourier transformation of $F(y)$ 
should be a scalar function of the relative momentum $q=p_Q - p_q = 
m_Q v + k - p_q$, 
\begin{equation}
	\widetilde{F}(q^2) = \int d^4 y e^{iq \cdot y} F(y) \, .
\end{equation}
In the heavy quark limit, the $k^2$- and $p_q^2$-dependences are 
suppressed by the heavy quark mass:
\begin{eqnarray}
	q^2 &=&  m_Q \Big(m_Q + 2 v \cdot (k-p_q) + O(1/m_Q) \Big) \,.
\end{eqnarray}
Also, due to heavy quark symmetry, $\widetilde{F}(q^2)$ must 
be a $m_Q$-independent function in the heavy quark 
limit. Besides, by momentum conservation [i.e. the $\delta$-function 
in Eq.(\ref{mhmf})], $k \sim p_H - p_q$ and hence $q^2 \sim (v \cdot p_H
-2v \cdot p_q)$. Since $\Psi$ describes the composite structure of 
heavy meson bound states in which $p_H=\overline{\Lambda}v$, then, 
without loss of the generality, we have replaced
$\widetilde{F}(q^2)$ by the $m_Q$-independent function $\Psi( v\cdot 
p_q)$ in Eq.~(\ref{mhmf}). 

A motivation of the above analysis is to provide a realistic field 
theory formulation of the intuitive picture of heavy hadrons; that 
is, in the heavy quark limit the heavy meson is a composite particle 
which can be described by a composite field operator of a heavy 
quark $h_v(k)$ coupled with the light degrees of freedom which is
sometimes called the ``brown muck" in the literature. 
From Eq.~(\ref{mhmf}), 
we see that it is natural to define the brown muck by 
\begin{equation}  \label{bmaq}
	\overline{q}_v(p_q) = \Psi(v \cdot p_q) 
		\overline{q}_0(p_q) \, .
\end{equation}
The above definition indicates that the brown muck consists of a 
light valence antiquark (which contains the tensoral structure of 
a spin-1/2 Dirac particle) coupled with a brown muck 
structure function $\Psi(v \cdot p_q)$, which effectively 
describes the dynamics of infinite number of $q\overline{q}$ 
pairs and gluons surrounding the light antiquark. Thus, according 
to Haag, Nishijima and Zimmermann \cite{HNZ}, we may rewrite the 
composite field operator of heavy mesons as a local operator: 
\begin{equation} \label{cfo}
	H_{ci}(x) = \overline{q}_v (x) \Gamma_i h_v (x) \, ,
\end{equation}
and its momentum representation is given by:
\begin{equation} \label{mhmf2}
	H_{ci}(v,p_H)=\int {d^4 k \over (2\pi)^4} {d^4 p_q \over
		 (2\pi)^4} (2\pi)^4 \delta^4 (p_H - k - p_q) 
		\overline{q}_v(p_q) \Gamma_i h_v(k) \, . 
\end{equation}

From Eq.(\ref{rshm}), the normalization
condition of the heavy meson bound states (with $p_H=\overline{\Lambda} 
v$) in the heavy quark limit is given by
\begin{equation} \label{hnc}
	\langle H_{ci'}(v') | H_{ci}(v) \rangle = 2(2\pi)^3 v^e \delta^3
		(\overline{\Lambda}v -\overline{\Lambda}v') \delta_{ii'} 
		\, ,
\end{equation}
where the superscript $e$ denotes the energy component in 
the momentum space, which can be either the $0$ or the $+$ component, 
depending on whether the light quark on-shall energy 
is picked on the light-front, $p^-_q=(p_{q\bot}^2 + m_q^2)/p_q^+$, 
or on the equal-time form, $p^0=\sqrt{\vec{p}_q^2 + m_q^2}$. 
Correspondingly, $\delta^3 (\overline{\Lambda}v-\overline{\Lambda}v')$ 
becomes $\delta(\overline{\Lambda}v^+-\overline{\Lambda}{v'}^+)\delta^2
(\overline{\Lambda}v_\bot-\overline{\Lambda}v'_\bot)$ or
$\delta^3(\overline{\Lambda}\vec{v} - \overline{\Lambda}\vec{v'})$.

Of course, a first-principles determination of the above composite
particle picture lies in the detailed form of $\Psi(v \cdot p_q)$.
How to solve $\Psi( v\cdot p_q)$ directly from QCD is one of the most 
interesting and difficult problems in strong interaction physics.
In the present paper, we shall treat 
$\Psi (v \cdot p_q)$ in a phenomenological manner.

\section{Effective field theory of heavy mesons}

Based on the above analysis of the composite particle picture of
heavy mesons in the heavy quark limit, we can now build an effective 
field theory to describe the heavy meson structure in HQET.

\subsection{Effective Lagrangian}
In principle, the heavy-meson composite particle structure is determined 
by the QCD Lagrangian for heavy and light quarks plus gluons, 
\begin{eqnarray}
	{\cal L} &=& {\cal L}_Q + {\cal L}_q  + {\cal L}_g \nonumber \\
		&=& \overline{Q}(i \! \not \! \! D - m_Q) Q + \overline{q}
		(i \! \not \! \! D - m_q) q - {1\over 4} F_a^{\mu \nu}
		F_{a\mu \nu} \, .
\end{eqnarray}
Using the $1/m_Q$ expansion to the heavy quark QCD Lagrangian 
\cite{Georgi90}, the above Lagrangian can be rewritten as
\begin{equation}
	{\cal L} = {\cal L}_0 + {\cal L}_{m_Q} \, ,
\end{equation}
where 
\begin{equation} \label{lo}
	{\cal L}_0 = \overline{h}_v i v \cdot D h_v + \overline{q}
		(i \! \not \! \! D - m_q) q - {1\over 4} F_a^{\mu \nu}
		F_{a\mu \nu}  \, 
\end{equation}
governs the structure of the heavy meson composite field operator of 
Eq.~(\ref{cfo}), and it has the heavy quark $SU_f(2) \otimes SU_s(2)$
flavor-spin symmetry (or simply the heavy quark symmetry, HQS), 
while, 
\begin{eqnarray}
	{\cal L}_{m_Q} &=&  \sum_{n=1}^\infty \Bigg({1 \over 2m_Q} \Bigg)^n 
		\overline{h}_v (i \! \not \! \! D_\bot)(-iv \cdot D)^{n-1}
		(i \! \not \! \! D_\bot) h_v \,   \nonumber \\
	&=& \sum_{n=1}^\infty \Bigg({1\over 2m_Q}\Bigg)^n {\cal L}_n
		\,  \label{l1m}
\end{eqnarray}
determines the $1/m_Q$ corrections to ${\cal L}_0$, which breaks the
heavy quark symmetry, where $D^\mu_\bot \equiv D^\mu - v^\mu v \cdot D$. 

To directly solve the structure of the composite field
operator of Eq.(\ref{cfo}) from ${\cal L}_0$ is not simple and may not
even be possible within the known framework of field theory. Instead
we shall introduce an effective Lagrangian to 
phenomenologically describe the heavy meson composite structure,
\begin{eqnarray}
	{\cal L}_0 \rightarrow {\cal L}^M_{\rm eff}(h_v,q,H_{ci})
	  &=& \overline{h}_v i v \cdot \partial h_v + \overline{q}_0(i \! 
		\not \! \partial - m_q) q_0 \nonumber \\
	  & & + P^\dagger_{v} \Big(iv \cdot \stackrel{\leftrightarrow}
		{\partial} - 2\overline{\Lambda}_0\Big) P_{v}  
	     - V^{\mu\dagger}_{v} \Big(iv \cdot \stackrel{\leftrightarrow}
		{\partial} - 2\overline{\Lambda}_0\Big) V_{v\mu}  \nonumber \\
	  & & + {G_0 } \Big(\overline{h}_v i\gamma_5 q_v P_{v} - 
		\overline{h}_v \gamma_\mu q_v V_{v}^\mu  + h.c. \Big) \,  ,
			\label{elcy}
\end{eqnarray}
which has the same heavy quark symmetry as ${\cal L}_0$,
where $\stackrel{\leftrightarrow}{\partial}\equiv\stackrel{\rightarrow}
{\partial}-\stackrel{\leftarrow}{\partial}$, $q_v$ is the brown muck 
field of the light degrees of freedom inside the heavy mesons, 
specified by Eq.~(\ref{bmaq}) in momentum space.  
$P_v$ and $V_v^\mu$ are the reduced pseudoscalar and vector heavy 
meson fields, respectively:
\begin{eqnarray}
	& & \Phi (x) = {1 \over \sqrt{M_P}} e^{-im_Q\,v \cdot x}
		P_v(x) \, , \\
	& & {\bf A}^\mu(x) = {1\over \sqrt{M_V}} e^{-i m_Q \, v \cdot x}
		V_v^\mu(x) \, ,  
\end{eqnarray}
with $v \cdot V_v = 0$, and $\overline{\Lambda}_0 \equiv M_0  - 
m_{Q_0}$ as the bare residual mass of the heavy 
mesons, where $M_0$ and $m_{Q_0}$ are defined here as the bare 
masses of heavy mesons and heavy quarks, 
respectively. The third and fourth terms in (\ref{elcy}) 
are directly obtained from the free Lagrangian of the pseudoscalar 
and vector fields with the above definition of the reduced 
fields,
\begin{equation}
   {\cal L}^M_{\rm free} = \Big(\partial^\mu \Phi^\dagger \partial_\mu
	\Phi - M^2_P |\Phi|^2\Big) - \Big({1\over 2} F^{\mu \nu \dagger}
	F_{\mu \nu} + M^2_V {\bf A}^{\mu \dagger}{\bf A}_\mu\Big) \, ,
\end{equation}
where $F^{\mu \nu} = \partial^\mu {\bf A}^\nu - \partial^\nu {\bf A}^\mu$.

In principle, the effective Lagrangian ${\cal L}^M_{\rm eff}$ may be 
derived from ${\cal L}_0$ by integrating 
out the gluon degrees of freedom in a nonperturbative way, which is 
unfortunately not practical at present. Therefore, we shall
assume that after integrating out the gluon degrees of freedom mediated 
between heavy and light quarks, we are led to
an effective Lagrangian of the form: 
\begin{eqnarray} 
	{\cal L}^{Qq}_{\rm eff} &=& \overline{h}_v i v \cdot \partial
		h_v + \overline{q}_0 (i \! \not \! \partial - m_q) q_0 
		\nonumber \\
	 & & + g_0^2 \Big(\overline{h}_v i\gamma_5 q_v \overline{q}_v
	  i\gamma_5 h_v - \overline{h}_v \gamma_\mu q_v \overline{q}_v
		\gamma^\mu h_v +\cdots	\Big) \,  ,  \label{el4q}
\end{eqnarray}
in which the complicated nonperturbative QCD dynamics to the heavy 
meson structure is effectively described by introducing the brown 
muck field $q_v$ through the universal structure function $\Psi(v 
\cdot p_q)$ of Eq.(\ref{bmaq}). {\it A consistency condition for 
the composite particle 
picture of heavy mesons in the heavy quark limit can be determined
by demanding the equivalence of Eqs.(\ref{elcy}) and (\ref{el4q}).}
Note that a priori the structure of $q_v(x)$ can be different in
Eqs.(\ref{elcy}) and (\ref{el4q}), however the consistency requirement
demands that they must be the same \cite{cheung97a}.

\subsection{Dynamical description of the composite particle 
	structure}

We can now dynamically describe the heavy mesons in the 
heavy quark limit as composite particles in the above two effective
field theories. 
In terms of Eq.~(\ref{el4q}), the composite particle structure of a 
heavy-light quark field $\overline{q}_v \Gamma_i h_v$ can be 
obtained by considering the heavy-light  quark scattering in the 
chain approximation, as shown in Fig.~1 \cite{Lurie}. 
In the pseudoscalar channel, apart from the factors come from the 
external quark lines, the scattering amplitude is given by 
\begin{equation}
	{\cal A}_{Qq} = g^2_0 \Psi(v \cdot p_q) \Psi^*(v \cdot p'_q)
		{i \over 1 - g^2_0 \Pi_H (v\cdot p_H)} \, , 
\end{equation}
where $p_H= k+ p_q$ is the momentum transfer, $k$ the residual 
momentum of the heavy quark, $p_q$ the momentum carried by the light quark,
and $\Pi_H(v \cdot p_H)$ the ``self-energy" correction
from the heavy-light quark loop depicted in Fig.~1(a), 
\begin{eqnarray}  
	-i \Pi_H (v \cdot p_H) &=& (-i)^2 (-1) \int {d^4 p_q 
		\over (2\pi)^4} |\Psi(v \cdot p_q)|^2 {\rm Tr}
		\Bigg[\Gamma_P i {1 + \! \not \! v \over 
		2(v \cdot p_H - v \cdot p_q + i \epsilon)} \nonumber \\
	& & ~~~~~~~~~~~~~~~~~~~~~~~~~ \times \Gamma_P i {-\! \not \! p_q 
		+ m_q \over p_q^2 - m_q^2 + i \epsilon }  \Bigg]  \, ,
\end{eqnarray}
where $\Gamma_P = i \gamma_5$. Thus, we have
\begin{eqnarray}
	\Pi_H (v \cdot p_H) &=& i \int {d^4 p_q \over (2\pi)^4} 
		{|\Psi(v \cdot p_q)|^2 \over (v \cdot p_H - v 
		\cdot p_q + i \epsilon)}\, { 2(v \cdot p_q + m_q) 
		\over (p_q^2 - m_q^2 + i \epsilon)} \, , \label{chcpg}
\end{eqnarray}

The existence of a stable heavy pseudoscalar composite meson with the
residual mass $\overline{\Lambda}$ implies 
that the above amplitude has a pole at $p_H=\overline{\Lambda} v$ 
(which corresponds to $P_H=M_H v$). This leads to 
\begin{equation}
	g^2_0 = {1\over \Pi_H( \overline{\Lambda}) } \, . 
\end{equation}
If we expand $\Pi_H(v \cdot p_H)$ by 
\begin{equation}  \label{see}
	\Pi_H(v \cdot p_H) = \Pi_H (\overline{\Lambda}) + (v \cdot p_H 
		- \overline{\Lambda}) \Pi'_H(\overline{\Lambda}) 
		+ \Pi^r_H(v \cdot p_H)  \, , 
\end{equation}
then, the amplitude can be 
expressed as 
\begin{equation}  \label{4qm}
	{\cal A}_{Qq} = -i{\Psi(v\cdot p_q)\Psi^*(v\cdot p'_q) \over 
        (v\cdot p_H - \overline{\Lambda})~ 
		\Pi'_H(\overline{\Lambda}) + \Pi^r_H(v \cdot p_H) } \, .
\end{equation}

On the other hand, in Eq.~(\ref{elcy}) the heavy meson field 
appears as a fundamental field. The physical (i.e., renormalized)
meson structure can be determined by considering a similar
heavy-light quark scattering process shown in Fig.~1(b). The amplitude is
\begin{equation}
	{\cal A}_{M}=G_0^2 \Psi(v \cdot p_q) \Psi^* (v \cdot p_q')
		\Delta_H (v \cdot p_H) \, , \end{equation}
where $\Delta (v \cdot p_H)$ is the heavy meson propagator:
\begin{equation}
	\Delta_H (v \cdot p_H) = {i \over 2 (v \cdot p_H - 
		\overline{\Lambda}_0) - G_0^2 \Pi_H (v \cdot p_H)} \, , 
\end{equation}
with $\Pi_H( v \cdot p_H)$ being given by Eq.~(\ref{chcpg}).
Using the expansion (\ref{see}),  the above amplitude can be
recast into
\begin{equation}  \label{ccm}
	{\cal A}_{M}=G ^2 \Psi(v \cdot p_q) \Psi^*(v \cdot p'_q)
		{i \over 2 (v \cdot p_H - \overline{\Lambda})- 
		G ^2 \Pi^r_H (v \cdot p_H) } \, ,
\end{equation}
with
\begin{eqnarray}   
	G  &=& Z_3^{1/2} G_0~~, ~~~ Z_3 = 1 + {G ^2 \over 2} \Pi'_H
		(\overline{\Lambda})  \label{rncd} \, , \\
	\overline{\Lambda} &=& \overline{\Lambda}_0 + {G_0^2\over 2} \Pi_H
		(\overline{\Lambda})  \, ,
\end{eqnarray}
where $Z_3$ is the wave function renormalization constant of the heavy 
meson field $P_v$.

In order that the pseudoscalar heavy meson, being a composite particle 
with the structure $\overline{q}_v \gamma_5 h_v$, can legitimately be 
represented by the field operator $P_v$, the two scattering amplitudes 
Eqs.~(\ref{ccm}) and (\ref{4qm}) must be the same. This results in 
\begin{equation}
	G ^{-2} = - {1\over 2} \Pi'_H (\overline{\Lambda})  
	  = i \int {d^4 p_q \over (2\pi)^4} {|\Psi(v \cdot p_q)|^2
		\over (\overline{\Lambda} - v \cdot p_q + i \epsilon)^2} 
		\,{ v \cdot p_q + m_q \over (p^2 - m_q^2 + i \epsilon)}  
		\label{wfrn} \, , 
\end{equation}
and
\begin{eqnarray}
	\overline{\Lambda}  &=& \overline{\Lambda}_0 + G_0^2
		\Pi_H (\overline{\Lambda} ) \nonumber \\
	   &=& \overline{\Lambda}_0 + i G_0^2\int {d^4 p_q \over 
		(2\pi)^4} {|\Psi(v \cdot p_q)|^2 \over 	(\overline{\Lambda} 
		- v \cdot p_q + i \epsilon)}\, { v \cdot p_q + m_q 
		\over (p^2 - m_q^2 + i \epsilon)}  \, . \label{massrn} 
\end{eqnarray}	 
We therefore obtain from Eq.~(\ref{rncd}) that 
\begin{equation}
	Z_3=0 \, . 
\end{equation}
The above 
results have a clear physical interpretation. The fact $Z_3=0$ 
implies that the bare fundamental field $P_{v} = Z_3^{1/2} P^R_v $ 
does not exist. In other words, the physical meson field $P^R_v$ 
must be a {\it composite particle operator}. 
Such a realization of the composite particle  
relies upon the introduction of
the composite particle structure function $\Psi(v \cdot p_q)$ in 
the effective Lagrangians (\ref{elcy}) and (\ref{el4q}). 
One can always choose a structure function $\Psi(v \cdot p_q)$ 
to ensure the finiteness of $\Pi_H (\overline{\Lambda})$ 
and $\Pi'_H(\overline{\Lambda})$. Without invoking such a 
structure function in the effective four-fermion point interaction, 
$\Pi'_H(0)$ is divergent and hence $G=0$, which leads to obvious 
inconsistency in the effective theory.  This shows the importance of 
$\Psi(v\cdot p_q)$ in our construction, which is not surprising 
since, as we will see shortly, $\Psi(v \cdot p_q)$ actually corresponds 
to a heavy meson wave function, while the renormalized coupling constant 
$G $ of Eq.~(\ref{wfrn}) is just the wave function 
normalization constant. 

Similar discussion for the vector meson structure can be easily 
carried out. In the heavy quark limit, the heavy quark symmetry 
of the effective theory ensures that the vector 
meson composite particle has the same structure as the pseudoscalar
particle. We will therefore not repeat the similar derivation here
for vector mesons.

\subsection{The effective field theory of heavy mesons with Feynman rules}

After determining the composite particle structure of heavy
mesons in the heavy quark limit, we can proceed to evaluate various heavy 
meson properties by combining the effective Lagrangian 
${\cal L}_{\rm eff}^{MR}$, which obeys heavy quark symmetry, with the
$1/m_Q$ corrections from ${\cal L}_{m_Q}$:
\begin{equation}
	{\cal L}_{\rm eff} = {\cal L}^{MR}_{\rm eff} + {\cal L}_{m_Q} \, ,
\end{equation}
where
\begin{eqnarray}
	{\cal L}^{MR}_{\rm eff}
	  &=& \overline{h}_v i v \cdot \partial h_v + \overline{q}(i \! 
		\not \! \partial - m_q) q \nonumber \\
	  & & + P^\dagger_{v} \Big(iv \cdot \stackrel{\leftrightarrow}
		{\partial} - 2\overline{\Lambda} \Big) P_{v}  
	     - V^{\mu\dagger}_{v} \Big(iv \cdot \stackrel{\leftrightarrow}
		{\partial} - 2\overline{\Lambda} \Big) V_{v\mu}  \nonumber \\
	  & & + G \Big(\overline{h}_v i\gamma_5 q_v P_{v} - 
		\overline{h}_v \gamma_\mu q_v V_{v}^\mu  + h.c. \Big) \,  ,
			\label{elcy1}  \\
	{\cal L}_{m_Q} &=&  \sum_{n=1}^\infty \Bigg({1 \over 2m_Q} \Bigg)^n 
		\overline{h}_v (i \! \not \! \! D_\bot)(-iv \cdot D)^{n-1}
		(i \! \not \! \! D_\bot) h_v \, ,  
\end{eqnarray}
$G$ is determined by Eq.~(\ref{wfrn}), 
$\overline{\Lambda}$ is the physical residual mass of heavy
mesons in the heavy quark limit, $M_H=m_Q + \overline{\Lambda}$,
$q_v = \Psi(v \cdot p_q) q$ and $\Psi(v \cdot p_q)$ describes
the heavy meson structure and is a phenomenological input at this
level. The Lagrangian ${\cal L}^{MR}_{\rm eff}$ gives the nonperturbative 
effective coupling of heavy mesons with heavy-light quarks. And
${\cal L}_{m_Q}$ is treated as a perturbation to ${\cal L}^{MR}_{\rm eff}$,
which contains all the $1/m_Q$ corrections. Then a practical 
evaluation scheme can be developed in terms of the standard Feynman 
diagrammatic rules:

(i) The heavy meson bound state ($p_H=\overline{\Lambda}v$) in the heavy 
quark limit gives a vertex as follows:
\begin{equation}
\begin{picture}(65,30)(0,38)
\put(0,37){$- - -$}
\put(33,40){\circle*{6} }
\put(33,41){\line(2,1){20}}
\put(33,40){\line(2,1){20}}
\put(33,39){\line(2,-1){20}}
\end{picture}
	: ~~~~~~~ -i G  \Psi^*(v \cdot p_q) \Gamma_H \, ,
\end{equation}
\begin{equation}
\begin{picture}(65,30)(0,38)
\put(20,37){$- - -$}
\put(20,40){\circle*{6} }
\put(20,41){\line(-2,1){20}}
\put(20,40){\line(-2,1){20}}
\put(20,39){\line(-2,-1){20}}
\end{picture}
	: ~~~~~~~ -i G  \Psi (v \cdot p_q) \Gamma_H \, ,
\end{equation}
with a momentum conservation factor $ (2\pi)^4 \delta^4 (\overline{
\Lambda}v-k-p_q)$, where $\Gamma_P = i \gamma_5$ for pseudoscalar
$(H=P)$, $ \Gamma_V = - \! \not \! \epsilon$ for vector mesons 
$(H=V)$, and $p_q$ is the momentum of the light degrees of freedom.
 
(ii) The internal line propagators for the heavy quark 
and the light antiquark are, 
\begin{eqnarray}
\begin{picture}(65,30)(0,38)
\put(0,40.5){\line(1,0){40}}
\put(19,40){\vector(1,0){2}}
\put(0,39.5){\line(1,0){40}}	
\put(19,28){$k$}
\end{picture}
	&:& ~~~~~~ i{\not \! v + 1 \over 2 (v \cdot k + i\epsilon)} \, ,\\ 
\begin{picture}(65,30)(0,38)
\put(0,40){\line(1,0){40}}
\put(21,40){\vector(-1,0){2}}	
\put(17.5,30){$-p_q$}
\end{picture}
 &:& ~~~~~~ i { - \!\not \! p_q + m_q \over p_q^2 - m_q^2 + i \epsilon} 
\, , \end{eqnarray} 
respectively, where $k$ is the residual momentum of the heavy quark,
and $m_q$ the constituent mass of the
light antiquark. 

(iii) For internal lines, integrate over the internal four-momenta,
\begin{equation}
	\int {d^4 k\over (2\pi)^4} ~~ {\rm and} ~~~ 
	\int {d^4 p_q\over (2\pi)^4} ~~\, , 
\end{equation}	
for heavy and light quarks, respectively. Also there is a factor
of $(-1)$ for each fermion loop. 

(iv) For all other vertices that do not attach to the bound states, 
the corresponding diagrammatic rules are standard from the the 
conventional field theory formulation. Most of these vertices mainly come
from ${\cal L}_{m_Q}$, hence the corresponding $1/m_Q$ corrections 
can be obtained from the standard perturbation field theory (see an 
explicit example in the next section). 

These are the Feynman rules needed for subsequent calculations in 
this effective field theory.

\section{Heavy meson properties in heavy quark limit}

In this section, we evaluate the basic heavy meson properties in 
the heavy quark limit within the present framework. These include the 
Isgur-Wise function, the decay constants and the axial-vector coupling 
constants of heavy mesons. 

\subsection{Isgur-Wise function}

In the heavy quark limit, the transition matrix elements of 
$B \rightarrow D$, and $B \rightarrow D^*$ are given by 
\begin{equation}
       \langle D (v') | \overline{h}_{v'}^c \Gamma
                h_v^b | B (v) \rangle ~~ {\rm and} ~~ 
		\langle D (v',\epsilon^*) |\overline{h}_{v'}^c 
		\Gamma h_v^b | B (v) \rangle ,     \label{5.3}
\end{equation}
where $\Gamma$ is a Dirac $\gamma$-matrix from the 
electroweak current of heavy quarks. Using the Feynman rules 
given in Sec.~IIIC, the hadronic matrix elements of $B \rightarrow 
D$ and $B \rightarrow D^*$ decays turn out to be (see Fig.~2(a)) 
\begin{eqnarray}
        & & \langle D (v') | \overline{h}_{v'}^c \Gamma      
                h_v^b | B (v) \rangle ={\rm Tr}\Big\{ \Gamma_P 
		\Big( {1+ \not{\! v}' \over 2} \Big)\Gamma \Big( 
		{1+\not{\! v} \over 2} \Big) \Gamma_P {\cal M} 
		\Big\}  \, , \label{ppd} \\
        & & \langle D (v',\epsilon^*) | \overline{h}_{v'}^c \Gamma 
		h_v^b | B (v) \rangle = {\rm Tr}\Big\{\Gamma_V^* 
		\Big( {1 + \not{\! v}' \over 2} \Big) \Gamma \Big( 
		{1+ \not{\! v} \over 2} \Big) \Gamma_P  {\cal M} \Big\} \, ,   
		\label{5.5} 
\end{eqnarray}
or
\begin{equation}
         \langle H' (v') | \overline{h'}_{v'} \Gamma 
		h_v | H (v) \rangle = {\rm Tr}\Big\{\Gamma_{H'} 
		\Big( {1 + \not{\! v}' \over 2} \Big) \Gamma \Big( 
		{1+ \not{\! v} \over 2} \Big) \Gamma_H  {\cal M} 
		\Big\} \, ,  \label{5.6}   
\end{equation}
for the general heavy meson decay process $H \rightarrow H'$,
where 
\begin{eqnarray} 
        {\cal M} = i G ^2 \int {d^4 p_q \over (2\pi)^4} {\Psi^*( v' \cdot 
		p_q ) \Psi ( v \cdot p_q ) \over (\overline{\Lambda} - 
		v \cdot p_q + i \epsilon) (\overline{\Lambda} - v \cdot p_q 
		+ i \epsilon)}{\not \! p_q - m_q \over p_q^2 - m_q^2 
		+ i \epsilon} \, , \label{bmk1} 
\end{eqnarray}
which is actually the transition matrix element of the light antiquark 
(brown muck).

Since ${\cal M}$ is fully covariant, it has the form 
\begin{equation}
	{\cal M} = A + B \not \! v + C \not \! v' \, , \label{bmk}
\end{equation}
where the coefficients $A, B, C, D$ can be easily determined to be
\begin{eqnarray}
	A &=& -i G ^2 \int {d^4 p_q \over (2\pi)^4} {\Psi^*(v' \cdot p_q ) 
		\Psi( v \cdot p_q) \over (p^2_q - m^2_q + i\epsilon)
		(\overline{\Lambda} - v \cdot p_q + i\epsilon) 
		(\overline{\Lambda} - v'\cdot p_q + i\epsilon) }~m_q \, , \\
	B &=& i G ^2 \int {d^4 p_q \over (2\pi)^4} {\Psi^*( v' \cdot p_q) 
		\Psi( v \cdot p_q) \over (p^2_q - m^2_q + i\epsilon)
		(\overline{\Lambda} - v \cdot p_q + i\epsilon) 
		(\overline{\Lambda} - v'\cdot p_q + i\epsilon) } \nonumber \\
	& & ~~~~~~~~~~~~~~~~~~~~~~~~ \times {1\over 2}
		\Bigg\{ {(v+v')\cdot p_q \over (1 + v \cdot v')} +
		{(v-v')\cdot p_q \over (1 - v \cdot v')} \Bigg\} \, , \\
	C &=& i G ^2 \int {d^4 p_q \over (2\pi)^4} {\Psi^*(v' \cdot p_q ) 
		\Psi( v \cdot p_q) \over (p^2_q - m^2_q + i\epsilon)
		(\overline{\Lambda} - v \cdot p_q + i\epsilon) 
		(\overline{\Lambda} - v'\cdot p_q + i\epsilon) } \nonumber \\
	& & ~~~~~~~~~~~~~~~~~~~~~~~~ \times {1\over 2}
		\Bigg\{ {(v+v')\cdot p_q \over (1 + v \cdot v')} -
		{(v-v')\cdot p_q \over (1 - v \cdot v')} \Bigg\} \, .
\end{eqnarray}
Then Eq.~(\ref{5.6}) can be simplified to
\begin{equation}
        \langle H' (v') | \overline{h'}_{v'} \Gamma      
                h_v | H (v) \rangle = -\xi(v \cdot v') {\rm Tr}\Big\{ 
		\Gamma_{H'} \Big( {1+ \not{\! v}' \over 2} \Big)\Gamma 
		\Big( {1+\not{\! v} \over 2} \Big) \Gamma_H 
		\Big\}  \, , \label{ppd1} 
\end{equation}
where $\xi(v \cdot v')$ is the Isgur-Wise function 
\begin{eqnarray}
	\xi (v \cdot v') &=& -(A - B - C)  \nonumber \\
	 &=& i G ^2 \int {d^4 p_q \over (2\pi)^4} \Psi^*(v' \cdot p_q) 
		\Psi (v \cdot p_q) \nonumber \\
	& & ~~~~~~~~~~~~~~~ \times {m_q + (v+v') \cdot p_q /(1+v \cdot v') 
		\over (p^2_q - m^2_q + i\epsilon)(\overline{\Lambda} - 
		v \cdot p_q + i\epsilon) (\overline{\Lambda} - v'\cdot 
		p_q + i\epsilon) } \, . \label{iwf}
\end{eqnarray}
At zero recoil $v \cdot v'=1$, i.e., $v'=v$, 
\begin{equation}
  \xi (1) = i G ^2 \int {d^4 p_q \over (2\pi)^4} |\Psi( v \cdot p_q )|^2 
		{ v \cdot p_q + m_q \over (p_q^2 - m_q^2 + i \epsilon)
		(\overline{\Lambda} - v \cdot p_q + i\epsilon)^2 } = 1 \, ,
\end{equation}
where we have used the result (\ref{wfrn}).  This is the well-known 
normalization of the Isgur-Wise function at the zero-recoil point,
as dictated by HQS.

\subsection{Decay constants in the heavy quark limit}

The decay constants of pseudoscalar and vector heavy mesons 
are defined by $\langle 0 | A^\mu | P(p) \rangle$ $= i f_P p^\mu$
and $\langle 0 | V^\mu | V(p,\epsilon) \rangle = f_V M_V \epsilon^\mu$
respectively, 
where $A^\mu = \overline{q}\gamma^\mu \gamma_5 Q$ and $V^\mu =
\overline{q}\gamma^\mu Q$. 
In the heavy quark limit, the meson decay constants are redefined as 
\begin{equation}
	\langle 0 | \overline{q} \gamma^\mu \gamma_5 h_v | P(v) 
		\rangle = i F_P v^\mu~~,
		~~~ \langle 0 | \overline{q}\gamma^\mu h_v | 
		V (v,\epsilon) \rangle = F_V \epsilon^\mu \, ,
\end{equation}
where in the heavy quark limit, $|P(p)\rangle=\sqrt{M_P}|P(v)\rangle$ 
and likewise for $|V(p,\epsilon)\rangle$, therefore,
\begin{equation}
	F_P = f_P \sqrt{M_P}~~,~~~ F_V = f_V \sqrt{M_V} \, .
\end{equation}
HQS demands that $F_V=F_P$.

Now, using the Feynman rules of the effective theory, it is very 
simple to evaluate the above matrix elements (see Fig.~2(b)): 
\begin{eqnarray}
	\langle 0 | \overline{q} \gamma^\mu \gamma_5 h_v |P (v) \rangle 
		&=& {\rm Tr}\Big\{ \gamma^\mu \gamma_5 {1 + \not \! v 
		\over 2} \Gamma_P {\cal M}_1 \Big\} \, , \\
	\langle 0 | \overline{q} \gamma^\mu h_v |V (v, \epsilon) \rangle 
		&=& {\rm Tr}\Big\{ \gamma^\mu {1 + \not \! v \over 2} 
		\Gamma_V  {\cal M}_1 \Big\} \, , 
\end{eqnarray}
where
\begin{eqnarray}  
	{\cal M}_1 &=& i G  \sqrt{N_c} \int {d^4 p_q \over (2\pi)^4} \Psi(v 
		\cdot p_q) { \not \! p_q - m_q \over (p_q^2 - m_q^2 
		+ i \epsilon)(\overline{\Lambda} - v \cdot p_q + 
		i\epsilon)} \nonumber \\
		&=& A_1 + B_1 \! \not \! v \, , \label{covint1}
\end{eqnarray}
and 
\begin{eqnarray}
	A_1 &=& - i G  \sqrt{N_c} \int {d^4 p_q \over (2\pi)^4} \Psi( v 
		\cdot p_q) {m_q \over (p_q^2 - m^2_q + i\epsilon)
		(\overline{\Lambda} - v \cdot p_q + i\epsilon)} \, ,  \\
	B_1 &=& i G  \sqrt{N_c} \int {d^4 p_q \over (2\pi)^4} \Psi(v 
		\cdot p_q) {v \cdot p_q \over (p_q^2 - m^2_q + i\epsilon)
		(\overline{\Lambda} - v \cdot p_q + i\epsilon) } \, .
\end{eqnarray}
Here $N_c=3$ is the number of colors. Thus, it is easily found:
\begin{eqnarray}
	F_P &=& - 2 (A_1 - B_1)  \nonumber \\
	&=& 2i G  \sqrt{N_c} \int {d^4 p_q \over 
		(2\pi)^4} \Psi(v \cdot p_q) { v \cdot p_q + m_q 
		\over (p^2_q - m_q^2 + i\epsilon)(\overline{\Lambda} 
		- v \cdot p_q + i\epsilon)}  \nonumber \\
	&=& F_V \, ,	\label{decc}
\end{eqnarray}
as expected from HQS. 

\subsection{Axial-vector coupling constants in the heavy quark limit}

The axial-vector coupling constants, $f$ and $g$, are defined by the 
transition matrix elements of $V \rightarrow P \pi$ and $ V \rightarrow 
V' \pi$ in the soft pion limit \cite{Yan94,cheng92}: 
\begin{equation}
         \langle P(v ) | A^a_\mu | V(v,\epsilon) \rangle ~~, ~~~~\langle
                V'(v,{\epsilon'}^*) | A^a_\mu | V(v,\epsilon) \rangle \, ,
\end{equation}
where $A^a_\mu = \overline{q} {\lambda^a \over 2} \gamma^\mu \gamma_5 q$ is 
the light quark axial-vector current. In the heavy quark limit,
\begin{eqnarray}
        & & \langle P(v) | q \cdot A^a | V(v,\epsilon) \rangle = i{f\over 2}
                \epsilon \cdot q  \, , \\
        & & \langle V'(v,{\epsilon'}^*) | q \cdot A^a | V(v,\epsilon)
                \rangle = ig \varepsilon^{\mu \nu \rho \sigma}
                q_\mu {\epsilon'}^*_\nu v_\rho \epsilon_\sigma  \, , 
\end{eqnarray}
where $q\simeq 0$ is the momentum carried by the soft-pion, and the SU(3) 
flavor matrix element $\chi^\dagger_{H'} \lambda^a \chi_H$ has been 
omitted in the above expressions. HQS predicts that $f=2g$ \cite{Yan94}. 

Diagrammatically, the above two matrix elements are represented by
Fig.~2c, from which one can straightforwardly write down these matrix 
elements: 
\begin{eqnarray}
        & & \langle P(v) | q \cdot A | V(v,\epsilon) \rangle
                = {1\over 2} {\rm Tr}\Big\{ \Gamma_P {1+\not \! v\over 2} 
		\Gamma_V {\cal M}_3 \Big\} \, , \nonumber \\
        & & \langle V'(v,{\epsilon'}^*) | q \cdot A | V(v,\epsilon)
                \rangle = {1\over 2} {\rm Tr} \Big\{\Gamma_V^* {1 +
                \not \! v\over 2} \Gamma_V {\cal M}_3 \Big\} \, ,
\end{eqnarray}
or simply
\begin{equation}  \label{sdc}
        \langle H'(v) | q \cdot A | H(v) \rangle ={1\over 2} {\rm Tr} 
	\Big\{ \Gamma_{H'} {1 + \not \! v\over 2} \Gamma_H {\cal M}_3 
		\Big\} \, ,
\end{equation}
where
\begin{eqnarray}
        {\cal M}_3 &=& -i G ^2 \int {d^4p_q \over (2\pi)^4} |\Psi(v \cdot 
		p_q)|^2 {(\not \! p_q - m_q) \not \! q \gamma_5 ( \not \! 
		p_q -m_q ) \over (p_q^2 - m_q^2 + i \epsilon)^2 
		(\overline{\Lambda} - v \cdot p_q + i\epsilon)} \nonumber \\
                &=& \Big(A_3 v \cdot q + B_3 \not \! q + C_3 v \cdot q
                \not \! v + D_3 \not \! q \not \! v \Big) \gamma_5 \, .
                \label{bmk2}
\end{eqnarray}
and the coefficients are given by
\begin{eqnarray}
        A_3 &=& i G ^2 \int {d^4p_q \over (2\pi)^4} |\Psi(v \cdot p_q)|^2 
        \,{ 2m_q v \cdot p_q \over (p_q^2 - m_q^2 + i \epsilon)^2 
		(\overline{\Lambda} - v \cdot p_q + i\epsilon)} \, , \\
        B_3 &=& -i G ^2 \int {d^4p_q \over (2\pi)^4} |\Psi(v \cdot p_q)|^2 
		{1\over 3}\,{2(v \cdot p_q)^2 + p_q^2 + 3 m_q^2 \over (p_q^2 
		- m_q^2 + i \epsilon)^2 (\overline{\Lambda} - v \cdot p_q + 
		i\epsilon)}  \, , \\
        C_3 &=& i G ^2 \int {d^4p_q \over (2\pi)^4} |\Psi(v \cdot p_q)|^2 
		{2\over 3}\, {4(v \cdot p_q)^2 - p_q^2 \over (p_q^2 - m_q^2 
		+ i \epsilon)^2 (\overline{\Lambda} - v \cdot p_q + 
		i\epsilon)} \, , \\
        D_3 &=& - A_3 \, .
\end{eqnarray}
Then, it is easily seen that Eq.(\ref{sdc}) can be simplified as
\begin{equation}
        \langle H'(v) | q \cdot A | H(v) \rangle = {g \over 2} {\rm Tr} 
		\Big\{\Gamma_{H'} {1 + \not \! v\over 2} \Gamma_H\not
          \! q\gamma_5 \Big\} \, ,
\end{equation}
with $f=2g$ and
\begin{eqnarray}
        g &=& -(B_3+D_3) \nonumber \\
	 &=& - i {G ^2\over 3} \int {d^4p_q \over (2\pi)^4} 
		|\Psi(v \cdot p_q)|^2 {(p_q^2-m_q^2) + 2(v \cdot p_q + m_q)
		(v \cdot p_q + 2m_q) \over (p_q^2 - m_q^2 + i \epsilon)^2 
		(\overline{\Lambda} - v \cdot p_q + i\epsilon)} \, . 
	\label{gg}
\end{eqnarray}

\section{The Determination of Heavy meson masses}

In this section, we determine the heavy meson masses up to order 
$1/m_Q$ within the effective theory.  

In the heavy quark limit, the heavy meson masses can be written as
$M_H = m_Q + \overline{\Lambda}$. In this limit, the pseudoscalar 
and vector mesons are degenerate. The correction to $M_M$ comes 
mainly from the leading HQS-breaking $1/m_Q$ corrections 
[see Eq.~(\ref{l1m})]: 
\begin{equation}
	{\cal L}_1 = \overline{h}_v (i D_\bot)^2 h_v + {g_s\over 2}
		\overline{h}_v \sigma_{\mu \nu}G^{\mu \nu} h_v 
		= {\cal O}_1 + {\cal O}_2 \, ,   \label{1ml}
\end{equation}
where $\sigma_{\mu \nu} = {i\over 2}[\gamma_\mu, \gamma_\nu]$ and
$G^{\mu \nu} = {i \over g_s}[D^\mu, D^\nu]$. 
With these $1/m_Q$ corrections included, the heavy meson masses can be 
expressed as \begin{eqnarray}  \label{massform}
	M_H = m_Q + \overline{\Lambda} - {1\over 2 m_Q}(\lambda_1
		+ d_H \lambda_2 ) \, , \label{mass}
\end{eqnarray}
where $\lambda_1$ comes from ${\cal O}_1$ and $d_H\lambda_2$ from
${\cal O}_2$ 
(see Fig.~3). $\overline{\Lambda}$, $\lambda_1$ and $\lambda_2$ are
the basic parameters in HQET. The parameter $\overline{\Lambda}$
is the residual mass of heavy mesons in the heavy quark
limit and is associated with the dynamical mass of the brown muck
\cite{Neubert94},
$\lambda_1$ parametrizes the common mass shift, 
and $\lambda_2$ describes the effect of the hyperfine 
mass splitting. In the rest frame of the heavy meson, the 
Clebsch-Gordon coefficient $d_H$ in Eq.~(\ref{mass}) 
is conventionally normalized in such a way that
\begin{eqnarray}
	d_H &=& -\langle H(v)|4\vec{S}_Q\cdot\vec{S}_\ell |H(v)\rangle   
		\nonumber\\
	&=& -2[S_{\rm tot}(S_{\rm tot}+1)-S_Q(S_Q+1)-S_\ell(S_\ell+1)],
		\label{cgc}
\end{eqnarray}
where $\vec{S}_Q~(\vec{S}_\ell)$ is the spin operator of the heavy 
quark (light degrees of freedom). Therefore, $d_H=-1$ for vector ($1^-)$ 
mesons, and $d_H=3$ for pseudoscalar ($0^-)$ mesons.

Using the Feynman diagrammatic rules of the effective theory, we can 
immediately write down the expressions for $\lambda_1$ and $d_H 
\lambda_2$ for heavy mesons as follows: 
\begin{eqnarray}
	\lambda_1 &=& \text{Tr} \Bigg\{\Gamma_H {1 + \not \! v \over 2}
		\Gamma_H {\cal M}_2 \Bigg\} \, , \\ 
	d_H \lambda_2 &=& {\rm Tr}\Bigg\{\Gamma_H {1+ \not \! v \over 2} 
		\sigma_{\mu \nu} {1 + \not \! v \over 2} \Gamma_H 
		{\cal M}_2^{\mu \nu} \Bigg\} \, , 
\end{eqnarray}
with
\begin{eqnarray}
 {\cal M}_2 &=& i G ^2\int {d^4 p_q\over (2\pi)^4} |\Psi( v\cdot p_q)|^2
		{(p_q^2 - (v \cdot p_q)^2)(\not \! p_q -m_q) \over
		(p_q^2 - m_q^2 + i \epsilon)(\overline{\Lambda} - v 
		\cdot p_{q} + i \epsilon)^2} \nonumber \\
	& & + i g_s^2C_fG ^2 \int {d^4 p_{1q} \over (2\pi)^4} 
		{d^4 p_{2q} \over (2\pi)^4} \Psi(v \cdot p_{1q})
		\Psi^*( v \cdot p_{2q}) \nonumber \\
	& & ~~~~~~~~~~~~~ \times { v \cdot (p_{1q} + p_{2q})v^\mu -(p_{1q}
		+p_{2q})^\mu \over (\overline{\Lambda} - v \cdot p_{1q}
		+ i \epsilon)( \overline{\Lambda} - v \cdot p_{2q} 
		+ i \epsilon)} \nonumber \\ 
	& & ~~~~~~~~~~~~~ \times D_{\mu \nu} {\not \! p_{2q} - m_q \over
		p_{2q}^2 - m_q^2 + i \epsilon} \gamma^{\nu} {\not \! 
		p_{1q} - m_q \over p_{1q}^2 - m_q^2 + i\epsilon} \, , 
		\label{lm1}\\
 {\cal M}_2^{\mu \nu} &=& g_s^2C_fG ^2 \int {d^4 p_{1q} \over (2\pi)^4} 
		{d^4 p_{2q} \over (2\pi)^4} \Psi(v \cdot p_{1q})
		\Psi^*( v \cdot p_{2q}) \nonumber \\
	& & ~~~~~~~~~~~~~ \times {(p_{1q} - p_{2q})^\mu \over (\overline{
		\Lambda} - v \cdot p_{1q} + i \epsilon )(\overline{\Lambda} 
		- v \cdot p_{2q} + i \epsilon)} \nonumber \\ 
	& & ~~~~~~~~~~~~~ \times D^{\nu \nu'} {\not \! p_{2q} - m_q \over
		p_{2q}^2 - m_q^2 + i \epsilon} \gamma_{\nu'} {\not \! 
		p_{1q} - m_q \over p_{1q}^2 - m_q^2 + i\epsilon} \, ,
		\label{lm2}
\end{eqnarray}
where $C_f$ is  a color factor, $C_f= {N_c^2-1 \over 2N_c} = 4/3$ for 
$N_c=3$, and $D^{\mu \nu}$ is the gluon propagator given by $D^{\mu 
\nu} = {-i g^{\mu \nu} \over q^2 +i\epsilon}$ in Feynman gauge
($q=p_{1q} - p_{2q}$). 

The mass shift parameter $\lambda_1$ can be further simplified to 
\begin{eqnarray}
	\lambda_1 &=& i G ^2\int {d^4 p_q\over (2\pi)^4} |\Psi( v\cdot 
		p_q)|^2 {2(p_q^2 - (v \cdot p_q)^2)(v \cdot p_q + m_q) 
		\over (p_q^2 - m_q^2 + i \epsilon)(\overline{\Lambda} 
		- v \cdot p_{q} + i \epsilon)^2} \nonumber \\
	& & - 2g_s^2C_fG ^2 \int {d^4 p_{1q} \over (2\pi)^4} 
		{d^4 p_{2q} \over (2\pi)^4} {\Psi(v \cdot p_{1q})
		\Psi^*( v \cdot p_{2q}) \over (\overline{\Lambda} - 
		v \cdot p_{1q} + i \epsilon)( \overline{\Lambda} - 
		v \cdot p_{2q} + i \epsilon)} \nonumber \\ 
	& & ~~~~~~~~~ \times { 1 \over ((p_{1q}-p_{2q})^2 + i\epsilon)
		(p_{1q}^2 - m_q^2 + i \epsilon)(p_{2q}^2 - m_q^2 + 
		i\epsilon)} \nonumber \\
	& & ~~~~~~~~~ \times \Big\{(v \cdot p_{1q} + m_q)[(p_{1q}+p_{2q})
		\cdot p_{2q} - v \cdot (p_{1q}+p_{2q}) v \cdot p_{2q}]
		\nonumber \\
	& & ~~~~~~~~~~~~~~+ (v \cdot p_{2q} + m_q)[(p_{1q}+p_{2q}) \cdot 
		p_{1q} - v \cdot (p_{1q}+p_{2q}) v \cdot p_{1q}] \Big\} \, .
		\label{lam1}
\end{eqnarray}
The hyperfine mass splitting parameter $\lambda_2$ can also be 
simplified. By using the identity 
\begin{equation}
	{1 + \not \! v \over 2} \sigma_{\mu \nu} {1 + \not \! v
		\over 2} v^\mu = 0  \, ,
\end{equation}
we find that only the antisymmetric component of ${\cal M}_2^{\mu\nu}$ 
contributes to $\lambda_2$. Thus we can let
\begin{equation}
	{\cal M}^{\mu \nu}_2 \rightarrow - {1\over 4} \sigma^{\mu \nu} 
		\zeta   \, .
\end{equation}
With the use of $\Gamma_H \not \! v=-\not\! v\Gamma_H$,
it is not difficult to find that $\zeta$ is given by
\begin{eqnarray}
	\zeta &=& {4\over 3}g_s^2C_fG ^2 \int {d^4 p_{1q} \over 
		(2\pi)^4}{d^4 p_{2q} \over (2\pi)^4} {\Psi(v \cdot p_{1q}) 
		\Psi^*( v\cdot p_{2q}) \over (\overline{\Lambda} - v 
		\cdot p_{1q}+ i \epsilon)( \overline{\Lambda} - v 
		\cdot p_{2q} + i \epsilon)} \nonumber \\
	& & ~~~~~~~~~~~~~~ \times { 1 \over ((p_{q1}-p_{2q})^2+i\epsilon)
		(p_{1q}^2 -m_q^2 + i\epsilon) (p_{2q}^2 - m_q^2 + i\epsilon)}
		\nonumber \\
	& & ~~~~~~~~~~~~~~\times \Big\{(v \cdot p_{1q} + m_q)[(p_{1q}-p_{2q}) 
		\cdot p_{2q} - v \cdot (p_{1q}-p_{2q}) v \cdot p_{2q}] 
		\nonumber \\
	& & ~~~~~~~~~~~~~~~~~~- (v \cdot p_{2q} + m_q)[(p_{1q}-p_{2q}) \cdot 
		p_{1q} - v \cdot (p_{1q}-p_{2q}) v \cdot p_{1q}] \Big\} \, . 
		\label{lam2}
\end{eqnarray}
Since
\begin{equation}
	{\rm Tr}\Bigg\{ \Gamma_H {1+ \not \! v \over 2} 
		\sigma_{\mu \nu} {1 + \not \! v \over 2} \Gamma_H 
		\sigma^{\mu \nu} \Bigg\} = \cases{ 
			-12~~~~~& for~$\Gamma_H = i\gamma_5$, \cr 
			4 ~~~~& for~$\Gamma_H = - \not \! \epsilon$, \cr}
\end{equation}
it follows that
\begin{equation} 
	d_H \lambda_2 = \zeta \times \cases{
			3~~~~~& ~{\rm for~pseudoscalar~mesons}, \cr 
			-1 &  ~{\rm for~vector~mesons}. \cr}
\end{equation}
Therefore,
\begin{equation}
	\lambda_2=\zeta \, .  \label{lam22}  
\end{equation}
The above results are valid in any Lorentz frame. In other words,
we obtain the result of Eq.~(\ref{cgc}) without setting the heavy 
meson in the rest frame. From 
Eq.~(\ref{massform}), we obtain the hyperfine mass splitting, 
\begin{equation}	\label{msp}
	\Delta M_{VP} = M_V - M_P = {2 \lambda_2 \over m_Q}  \, .
\end{equation}

As we can see from Eq.~(\ref{lam1}), $\lambda_1$ consists of two 
contributions: one is the kinetic energy of heavy quarks [the first 
term in Eq.~(\ref{lam1})], and the other is the effect of the 
chromo-electric interaction between the heavy quark and light degrees 
of freedom. For convenience, we redefine
\begin{equation}
	\lambda_1 \equiv -\langle \vec{k}^2 \rangle + \alpha_s 
		\overline{\lambda}_1~~, ~~~
	\lambda_2 \equiv \alpha_s \overline{\lambda}_2 \, , \label{repara}
\end{equation}
where $\alpha_s= g_s^2/4\pi$, and $\langle \vec{k}^2 \rangle$, 
$\overline{\lambda}_1$ and $\overline{\lambda}_2$ are denoted 
as the kinetic energy, the chromo-electric and chromo-magnetic
interaction parameters, respectively. These parameters depend
only on the structure function $\Psi(v \cdot p_q)$. 

\section{Comparison of the effective theory with the covariant 
	Light-front Model}

In the previous sections we have developed an effective 
field theory to describe heavy mesons as composite particles, 
and systematically evaluated the heavy meson properties in 
the heavy quark limit and the basic HQET parameters.  
In this section, we will compare  
the effective theory with the covariant light-front 
model of heavy mesons that we have constructed recently\cite{cheng97b}, 
and show the source of difficulty for developing a fully covariant 
reformulation of the currently used light-front quark model in the 
literature.

\subsection{Covariant light-front quark model as a special case}

So far we have not specified the form of the heavy meson structure 
function $\Psi(v \cdot p_q)$.  From the constraint that 
it forbids on-mass-shell decay of the heavy meson into $Q\bar q$,
a possible form is given by 
\begin{equation}
\Psi(v \cdot p_q)= (\overline\Lambda - v \cdot p_q) 
                   \varphi(v \cdot p_q)\label{Psi},
\end{equation}
where the function $\varphi(v \cdot p_q)$ is regular at 
$v \cdot p_q = \overline\Lambda$.  We shall further assume that 
(1) $\varphi(v \cdot p_q)$ is analytic except for isolated singularities 
when continued into the complex plane, and (2) it vanishes ``fast enough" 
as $|v \cdot p_q|\rightarrow\infty$. 
These two conditions allow us to perform the $dp^e-$ integrations in the 
expressions derived earlier by Cauchy's Theorem.  
Thus by closing the contours in the lower-half 
complex $p^e-$plane, we obtain 
\begin{eqnarray}
	\xi (v \cdot v') &=& i G^2 \int {d^4 p_q \over (2\pi)^4} \varphi^*(v' 
		\cdot p_q) \varphi (v \cdot p_q) \nonumber \\
	& & ~~~~~~~~~\times {m_q + (v+v') \cdot 
		p_q /(1+v \cdot v') \over (p^2_q - m^2_q + i\epsilon)
		 } \nonumber \\ 
	 &=& G^2 \int {d^4 p_q \over (2\pi)^4} (2\pi) \delta(p_q^2-m_q^2) 
		\theta(p_q^e) \varphi^*(v'\cdot p_q) \varphi 
		(v \cdot p_q) \nonumber \\
	& & ~~~~~~~~~\times \Big[{m_q 
		+ (v+v') \cdot p_q /(1+v \cdot v')}\Big]\, , \label{iwf1}
\end{eqnarray}
\begin{eqnarray}
	F_P &=& 2 i G^2\sqrt{N_c} \int {d^4 p_q \over (2\pi)^4} \varphi(v 
		\cdot p_q) {{ v \cdot p_q + m_q } \over p_q^2
		- m_q^2 + i \epsilon} \nonumber \\  
	  &=& 2 G^2\sqrt{N_c} \int  {d^4 p_q \over (2\pi)^4} (2\pi) 
		\delta(p_q^2 -m_q^2)\theta (p_q^e) \varphi(v \cdot p_q) 
		{( v \cdot p_q + m_q)}  \, ,	\label{decc1}
\end{eqnarray}
and
\begin{equation}
	G ^{-2} = \int {d^4 p_q \over (2\pi)^4} (2\pi) \delta (p^2_q
		 - m_q^2) \theta(p^e) (v \cdot p_q+m_q)  
                 |\varphi (v \cdot p_q)|^2.  \label{G2}
\end{equation}

Now if we take 
\begin{equation}
\varphi(v\cdot p_q)={\Phi_{LF}(v\cdot p_q)\over \sqrt{v\cdot p_q+m_q}},
\label{varphi1}
\end{equation}
where $\Phi_{LF}(v\cdot p_q)$ is a normalized light-front wave 
function, and let $p^e=p^-$, then these results are exactly the same as 
those obtained in the covariant light-front quark model\cite{cheng97b}.
Thus we see that, for these quantities, the covariant light-front
quark model results can be recovered here with a special choice of 
$\varphi(v\cdot p_q)$.

\subsection{Beyond the covariant light-front quark model}

Next, we shall consider the mass parameters $\lambda_1$ and $\lambda_2$ 
(see Sec.~V). 
As before, using Eq.~(\ref{Psi}) and 
after performing the $p^e$-integrals, we obtain 
\begin{eqnarray}
	\lambda_1 &=& 2 G^2 \int {d^4 p_q\over (2\pi)^4} (2\pi)\delta(p_q^2
		-m_q^2) \theta (p_q^e) |\varphi ( v\cdot p_q)|^2 
		[m_q^2 - (v \cdot p_q)^2] (v\cdot p_q+m_q)\nonumber \\
	& & + 2G^2g_s^2C_f \int {d^4 p_{1q} \over (2\pi)^4} {d^4 p_{2q} 
		\over (2\pi)^4} {\varphi(v \cdot p_{1q})\varphi^*( v 
		\cdot p_{2q}) }  \nonumber \\
	& & ~~~~~~~~~ \times \Bigg\{ {(2\pi)\delta(p_{1q}^2-m_q^2) 
		\theta (p_{1q}^e) (2\pi)\delta(p_{2q}^2 -m_q^2) 
		\theta (p_{2q}^e) \over (p_{1q}-p_{2q})^2 + i\epsilon} 
		\nonumber \\ 
	& & ~~~~~~~~~ + {(2\pi)\delta(p_{1q}^2-m_q^2) \theta 
		(p_{1q}^e) (2\pi)\delta((p_{1q}-p_{2q})^2) 
		\theta (p_{2q}^e-p_{1q}^e) \over p_{2q}^2- m_q^2} 
		\nonumber \\ 
	& & ~~~~~~~~~ + {(2\pi)\delta(p_{2q}^2-m_q^2) \theta 
		(p_{2q}^e) (2\pi)\delta((p_{1q}-p_{2q})^2) 
		\theta (p_{1q}^e-p_{2q}^e) \over p_{1q}^2-m_q^2}
		\Bigg\} \nonumber \\ 
	& & ~~~~~~~~~\times \Big\{(v \cdot p_{1q} + m_q)[(p_{1q}+p_{2q}) 
		\cdot p_{2q} - v\cdot (p_{1q}+p_{2q}) v \cdot p_{2q}]
		\nonumber \\
	& & ~~~~~~~~~~~~~~+ (v \cdot p_{2q} + m_q)[(p_{1q}+p_{2q})\cdot 
		p_{1q}- v \cdot(p_{1q}+p_{2q}) v \cdot p_{1q}] \Big\} \, ,
		\label{lam1l1} \\
	\lambda_2 &=& {4\over 3}G^2g_s^2C_f \int {d^4 p_{1q} \over (2\pi)^4} 
		{d^4 p_{2q} \over (2\pi)^4} {\varphi(v \cdot p_{1q}) 
		\varphi^*( v\cdot p_{2q}) } \nonumber \\
	& & ~~~~~~~~~ \times \Bigg\{ {(2\pi)\delta(p_{1q}^2-m_q^2) 
		\theta (p_{1q}^e) (2\pi)\delta(p_{2q}^2 -m_q^2) 
		\theta (p_{2q}^e) \over (p_{1q}-p_{2q})^2 + i\epsilon} 
		\nonumber \\ 
	& & ~~~~~~~~~ + {(2\pi)\delta(p_{1q}^2-m_q^2) \theta 
		(p_{1q}^e) (2\pi)\delta((p_{1q}-p_{2q})^2) 
		\theta (p_{2q}^e-p_{1q}^e) \over p_{2q}^2- m_q^2} 
		\nonumber \\ 
	& & ~~~~~~~~~ + {(2\pi)\delta(p_{2q}^2-m_q^2) \theta 
		(p_{2q}^e) (2\pi)\delta((p_{1q}-p_{2q})^2) 
		\theta (p_{1q}^e-p_{2q}^e) \over p_{1q}^2-m_q^2}
		\Bigg\} \nonumber \\ 
	& & ~~~~~~~~~\times \Big\{(v \cdot p_{1q} + m_q)[(p_{1q}-p_{2q}) 
		\cdot p_{2q}- v \cdot (p_{1q}-p_{2q}) v \cdot p_{2q}]
		\nonumber \\
	& & ~~~~~~~~~~~~~~- (v \cdot (p_{2q} + m_q)[(p_{1q}-p_{2q}) \cdot 
		p_{1q}- v \cdot (p_{1q}-p_{2q}) v \cdot p_{1q}] \Big\} \, .
		\label{lam2l1}
\end{eqnarray}

%
In each of the above expressions, the first delta-function term 
inside the bracket comes from
on-mass-shell light antiquarks.  With $\varphi(v\cdot p_q)$ given by 
Eq. (\ref{varphi1}), one can easily check that this contribution is exactly 
what one 
would  obtain in the covariant light-front quark model \cite{cheng97b} 
(where the expression for $\lambda_1$ is not explicitly displayed).  
The rest of the terms 
are due to the off-mass-shell contributions of 
the antiquarks, they correspond to the so-called $Z$-diagram 
contributions which cannot be obtained in any conventional quark model 
formulations. 

As another example, we examine the axial-vector coupling constant $g$, 
which involves a purely light-quark current. 
To obtain a covariant result, one 
has to keep all the time-ordered (including the so-called 
$Z$-diagram) contributions, which is obviously beyond the scope of the 
light-front quark model \cite{cheung97}. However this is not a problem  
in the present framework, since Feynman diagrams naturally 
contain all possible time-orderings. 
As a result, covariance is automatically preserved in deriving all the 
results in the previous sections.

Now substituting Eq. (\ref{Psi}) into Eq. (\ref{gg}), and integrating 
over $p^-$, we obtain 
\begin{eqnarray}
       g &=& {G^2\over 3} \int {dp_q^+dp_{q\perp} \over (2\pi)^3 2 p^+}
	 \Bigg\{ |\varphi(v \cdot 
	p_q)|^2 {(v \cdot p_q + 2m_q)(v\cdot p_q+m_q)\over X}
	- {(\overline{\Lambda}- v \cdot p_q)\over X} \nonumber \\
	&&\times \Bigg[{2 v\cdot p_q + 3 m_q +X} +
	(v \cdot p_q + 2m_q)(v\cdot p_q+m_q) {\partial \over \partial 
	(v \cdot p_q)}\Bigg]  |\varphi (v \cdot p_q)|^2  
		\Bigg\} \, , \label{lfg} 
\end{eqnarray}
where $X=p_q^+/v^+$ and $p_q^2=m_q^2$.  
Obviously, this expression cannot be obtained from any type of 
light-front quark models because it involves a derivative of the wave 
function, which is very unusual and cannot be simply related to some 
matrix elements of hadronic bound states in quark models.  
However it is interesting to note that  
the first term in Eq.~(\ref{lfg}) 
is what one would get for $g$ if we naively calculate it in the 
light-front quark model \cite{cheng97b}.  Therefore 
the second term in (\ref{lfg}) should be the $Z$-diagram 
contribution not present in the light-front quark model formulation. 

For completeness, we also display the resulting expression for $g$  
by integrating over $p^0$ in Eq.~(\ref{gg}): 
\begin{eqnarray}
       g &=& {G^2\over 3} \int {dp^3_q\over (2\pi)^3 2 p^0}
	 \Bigg\{ |\varphi(v \cdot p_q)|^2 {(v \cdot p_q + 2m_q)
	(v\cdot p_q+m_q)\over X}- {(\overline{\Lambda}- v \cdot 
	p_q)\over X} \Bigg[2 v\cdot p_q +3 m_q \nonumber \\
	&& + X + (v \cdot p_q + 2m_q)(v\cdot p_q+m_q) 
	\Big(-{1\over p^0}+ {\partial \over \partial (v \cdot p_q)}\Big)
	\Bigg]  |\varphi (v \cdot p_q)|^2  \Bigg\} \, , \label{etg}
\end{eqnarray}
where $X=p_q^0/v^0$ and $p_q^2=m_q^2$.  

Thus we have shown that, by the specific choice of Eq.~(\ref{varphi1}), 
our effective field theory can 
reproduces the covariant light-front quark model \cite{cheng97b} 
results for those quantities ($\xi(v\cdot v'), F_P,$ and $F_V$) 
which do not have $Z$-diagram contributions.  
For other quantities ($\lambda_1, \lambda_2,$ and $g$), the extra 
$Z$-diagram contributions missed in the 
light-front quark model can be explicitly identified here. 
Hence the effective field theory presented here 
has the simplicity of a conventional quark model, and is fully covariant 
at the same time. 

\subsection{Choice of $\Psi(v \cdot p_q)$} 

Now an important question is what kind of $\Psi(v\cdot p_q)$ 
will allow a proper analytic continuation into the complex plane, so that 
the $p^e$-integrals can be evaluated by Cauchy's Theorem,  and 
comparisons can be made with results obtained in the quark model. 
In the light-front formulation, there are 
three type of mesonic wave functions which have been widely 
used in the literature. One of them, the so-called BSW wave 
function (or BSW model) has already been ruled out since it 
explicitly breaks the relativistic covariance and thereby leads to 
inconsistent results in HQET, as we have shown in recent
publications \cite{cheng97a,cheng97b}. The other two wave functions,
namely, the Gaussian-type wave function and the invariant light-front
mass wave function, have the following forms in the
heavy quark limit \cite{cheng97b}:
\begin{eqnarray} 
	\Phi_{LF}^G (v \cdot p_q) &=& {\cal N}_G \sqrt{v \cdot p_q} 
		\exp \Bigg\{ -{1\over 2\omega^2} \Big[(v\cdot p_q)^2 
		-m_q^2 \Big] \Bigg\} \, ,  \label{gaus}  \\
	\Phi_{LF}^M (v \cdot p_q) &=& {\cal N}_M \exp \Bigg\{ 
		- {v \cdot p_q \over \omega} \Bigg\} \, . \label{wfus} 
\end{eqnarray}
These two wave functions apparently have a covariant form
\cite{cheng97b}. However, in the light-front quark model 
formulation, the light antiquark is always on the mass shell: 
$p^-_q = {p_{q\bot}^2 +m_q^2) \over p_q^+}$.  
It is not obvious how to extend these wave functions to be used in a 
4-dimensional covariant framework.  
If $p_q$ were allowed to go off-shell in these wave functions, 
then $\Phi^M_{LF}(v\cdot p_q)$ would be unbounded when  
$|p^e_q|\rightarrow\infty$, whereas $\Phi^G_{LF}(v\cdot p_q)$ would  
also be unbounded when analytically continued into to the complex plane. 
To fix these problems, one could instead use $|v\cdot p_q|$ in the above 
expressions, but then both wave functions would not be analytic in the 
complex plane.
In any case, we find that it is not possible to modify the exponential 
form so that the two conditions stated below Eq. (\ref{Psi}) are satisfied.


Besides the exponential-type functions, 
we can also consider the so-called Lorentzian-type dependence 
for $\Psi(v\cdot p_q)$.  In the heavy quark limit, we may 
write $\Psi (v\cdot p_q)$ as 
\begin{equation}  \label{lwf}
	\Psi_n (v \cdot p_q) = {\overline{\Lambda} - v\cdot p_q
		 \over (v \cdot p_q + \omega - i\epsilon)^n} \, ,
\end{equation}
where an $-i\epsilon$ has been inserted in the denominator so that 
$\Psi_n(v\cdot p_q)$ is analytic in the lower half complex-$p^e$ plane.
Consequently by closing the integration contours in the lower half 
complex-$p^e$ plane, we will not pick up contributions from the poles of 
$\Psi_n(v\cdot p_q)$.
It is easy to check that this choice of $\Psi(v\cdot p_q)$ 
corresponds to a covariant light-front wave function of the form
\begin{equation}
	\Phi^{n}_{LF} (v \cdot p_q) = {\cal N}_L {\sqrt{ v\cdot p_q
		+ m_q} \over (v \cdot p_q + \omega )^n  } ~~~
		~~ n > 2 \, ,
\end{equation}
where ${\cal N}_L$ is a normalization constant, and $p_q^2 = m_q^2$. 
Here we require that $n > 2$ to ensure the finiteness of all the integrals.

In the following, we shall present some numerical analyses to further
examine the predictive power of the effective theory with the choice 
Eq.~(\ref{lwf}) for $\Psi(v \cdot p_q)$.

\section{Numerical Calculation and Discussions}

\subsection{Short-distance corrections}

Before performing the numerical computation, it should be stressed 
that all the quantities we have evaluated in the effective field theory
are mainly governed by the long-distance physics of HQET.  That is, 
a renormalization scale $\mu$ ($\Lambda_{QCD} << \mu << m_Q$) has been
implicitly employed in the structure function $\Psi(v \cdot p_q)$.  
It is necessary to take into account short-distance QCD corrections 
to match with the full QCD description:
\begin{equation}
	\langle {\cal O} \rangle_{\rm QCD} = C_0(\mu) \langle {\cal O}_0
		(\mu) \rangle_{\rm HQET} + {C_1(\mu) \over 2m_Q(\mu)} 
		\langle {\cal O}_1 (\mu) \rangle_{\rm HQET}  + \cdots  \, ,
\end{equation}
where the Wilson coefficients $C_i(\mu)$, which account for the 
short-distance corrections, have been computed in the literature
\cite{Neubert94}.

Explicitly, the $B-$meson decay constant $f_B$ is given by
\begin{equation}  \label{fbdc}
	f_B = {1\over \sqrt{M_H}}\Bigg[{\alpha_s(m_b) \over \alpha_s 
		(\mu)} \Bigg]^{-6/(33-2N_f)} F_P (\mu) \, ,
\end{equation}
where $F_P (\mu)$ is the decay constant defined in Sec.~IVB, 
and $N_f=4$ is the number of quark flavors with mass less than $m_b$. 

The Isgur-Wise function discussed in Sec.~IV.A
is also defined at the renormalization scale $\mu$. Its 
relation with the $\mu$-independent physical observable, for example, the
form factor $f_+(q^2)$ [or $F_1(q^2)]$ is given by \cite{Falk}
\begin{equation}
	f_+ (q^2) = {M_B+M_D\over 2\sqrt{M_B M_D}} \Bigg[{\alpha_s(m_b) 
                 \over \alpha_s 
		(m_c)} \Bigg]^{-6/25} \Bigg[{\alpha_s(m_c) \over \alpha_s 
		(\mu)} \Bigg]^{a_L(v\cdot v')} \xi(v \cdot v',\mu) \, ,
\end{equation}
where
\begin{eqnarray}
a_L(\omega) &=& {8\over 27}\,[\omega r(\omega)-1],   \nonumber \\
r(\omega) &=& {1\over \sqrt{\omega^2-1}}\,\ln(\omega+\sqrt{\omega^2-1}).
\end{eqnarray}
The axial-vector coupling constant $g$ does not receive short-distance 
corrections since it involves a light-quark current which is partially 
conserved.

Although the physical heavy meson mass is scale independent, individual 
contributions to it are modified by short-distance corrections.
Thus Eq.~(\ref{mass}), written separately for pseudoscalar and vector 
mesons, becomes 
\begin{eqnarray}
	M_H &=& m_Q(\mu) + \overline{\Lambda}(\mu) - {1\over 2m_Q(\mu)}
		\Big[\lambda_1(\mu) + 3 a_2(\mu) \lambda_2 (\mu) \Big]
		\, , \label{mass1} \\
	M_{H^*} &=& m_Q(\mu) + \overline{\Lambda}(\mu) - {1\over 2m_Q
		(\mu)}\Big[\lambda_1(\mu)-a_2(\mu) \lambda_2(\mu)\Big] \, ,
		\label{mass2}
\end{eqnarray}
where 
\begin{equation}
	a_2(\mu) = \Bigg[{\alpha_s (m_Q) \over \alpha_s (\mu)}
		\Bigg]^{-9/(33-2N_f)}  \, .
\end{equation} 
Note that all the basic nonperturbative parameters in HQET, i.e. $m_b(\mu)$, 
$\overline{\Lambda}(\mu)$, $\lambda_1(\mu)$ and $\lambda_2 (\mu)$, 
are $\mu$-dependent; only the physical masses are independent of the 
renormalization scale $\mu$.

It should be stressed that in the above expressions, we have distinguished
the heavy quark mass in the $1/m_Q$ expansion which is the running
heavy quark mass defined at the scale $\mu$, and 
the mass scale in the strong coupling constant $\alpha_s(m_Q)$ set at 
the heavy quark pole mass: $\alpha_s(m_Q^{\rm pole})$. In the literature, 
there are many discussions on the ambiguities in the definition of the
heavy quark mass due to the presence of 
the so-called renormalons \cite{BSUV}.
We believe that the above distinction of heavy quark masses in our 
formulation is theoretically consistent.

Using Eq.(\ref{repara}), Eqs.~(\ref{mass1}-\ref{mass2}) can be 
rewritten as 
\begin{eqnarray}
	M_H &=& m_Q(\mu) + \overline{\Lambda}(\mu) + {1\over 2m_Q(\mu)}
		\Big\{ \langle\vec{k}^2\rangle - \alpha_s(\mu) \Big[
		\overline{\lambda}_1+3 a_2(\mu) \overline{\lambda}_2
		\Big]\Big\} \, , \label{mass11} \\
	M_{H^*} &=& m_Q(\mu) + \overline{\Lambda}(\mu) + {1\over 2m_Q(\mu)}
		\Big\{ \langle\vec{k}^2\rangle - \alpha_s(\mu)\Big[
		\overline{\lambda}_1 - a_2(\mu) \overline{\lambda}_2\Big]
		\Big\} \, .\label{mass22} 
\end{eqnarray}
Note that  $\langle \vec{k}^2 \rangle,~\overline{\lambda}_1$ and
$\overline{\lambda}_2$ depend only on the structure function
$\Psi(v \cdot p_q)$, and therefore also on the scale $\mu$. 
The heavy meson mass splitting between 
pseudoscalar and vector mesons is simply given by 
\begin{eqnarray}
	M_{H^*} - M_H = 2\, {\alpha_s(\mu) a_2(\mu)\over m_Q(\mu)}  
		\,\overline{\lambda}_2 \, .  \label{pbms}
\end{eqnarray} 

\subsection{Numerical analysis}

For the sake of demonstration, we first set 
$\mu=m^{\rm pole}_b \simeq 4.89$ GeV (the pole mass) \cite{mass}, 
and see if a consistent set of parameters 
can be found.  At this scale, short-distance corrections are not present. 
The light quark mass at this scale is given by the current quark
mass which is about a few MeV for up and down quarks. We take
$m_{u,d}(m_b^{\rm pole}) \simeq 5$ MeV. Then the parameter $\omega$ 
in the structure functions can be fixed by the decay constant $f_B$ 
via Eq.~(\ref{decc}). Taking 
\begin{equation}  \label{para1}
	f_B = 0.180 ~~ {\rm GeV},
\end{equation}
the values of $\omega$ for different structure functions $\Psi_n (v\cdot 
p_q)$ are listed in Table I. 
Since the strong coupling constant $\alpha_s$ is also known at 
$\mu=m^{\rm pole}_b$, knowing $\omega$, we can predict various heavy 
meson properties obtained in the effective theory at this scale. 

We first calculate the vector-pseudoscalar $B$ meson mass
difference which is determined by $\lambda_2$. As $\mu = 
m^{\rm pole}_b \gg \Lambda_{\rm QCD}$, $\alpha_s(m^{\rm pole}_b)$ 
can be determined from $\alpha_s (M_Z)$ by the perturbative evolution 
equation (at the one-loop level in the $\overline{\rm MS}$ scheme), 
\begin{equation}
	\overline{\alpha}_s (m^{\rm pole}_b) = {\alpha_s (M_Z) \over 1 + 
		\alpha_s (M_Z) \beta_0 \ln [(m^{\rm pole}_b)^2/ M^2_Z]
		/(4\pi) } \simeq 0.22 \, , 
\end{equation}
where $\beta_0 = 11-{2\over 3}N_f=11-8/3$ for $N_f=4$, $M_Z=91.19$ GeV,
and $\alpha_s (M_Z) = 0.119$ from experimental fits. Thus, the 
$B^* -B$ mass splitting at the scale $\mu = m_b^{\rm pole}$ in our
calculation is given by 
\begin{eqnarray}
  \Delta M_{B^*B}=M_{B^*} - M_B = 2\,{\overline{\alpha}_s(m^{\rm pole}_b) 
  \over  \overline{m}_b(m^{\rm pole}_b)} \, \overline{\lambda}_2 \, ,
\end{eqnarray}
with the $\overline{\rm MS}$ $b$-quark mass at this scale: 
$\overline{m}_b(m^{\rm pole}_b)= 4.339$ GeV \cite{mass}. The 
result is also listed in Table I, from which it is evident  
our results do not fit the experimental value of $\Delta 
M_{B^*B}=0.046$ GeV \cite{PDG}. This numerical analysis indicates 
that the real scale $\mu$ implicitly used in the effective theory should 
be lower than $m^{\rm pole}_b$. 

\vskip 0.5cm
{\footnotesize
Table I. The parameter $\omega$ in the structure functions $\Psi_n
(v\cdot p_q)$ [(\ref{lwf})] with different $n$ values fitted to $f_B
=180$ GeV at the scale $\mu=m_b^{\rm pole}$ and the resulting $\Delta 
M_{B^*B}$ and $\langle \vec{k}^2 \rangle$. With $n$ larger than 12, 
the change of $\Delta M_{B^*B}$ and $\langle \vec{k}^2 \rangle$ becomes 
insignificant.
\begin{center}
\begin{tabular}{|c|c|c|c|c|c|c|c|c|c|c|}  
\hline\hline n 		& 3 	& 4 	& 5 	& 6 	& 7 
	& 8 	& 9 	& 10 	& 11	& 12 \\ \hline
$\omega$ (GeV) 		& 0.100 & 0.374 & 0.689 & 1.011 & 1.335   
	& 1.660 & 1.986 & 2.312 & 2.638	& 2.965 \\ \hline
$\Delta M_{B^*B}$ (GeV) 	& 0.011 & 0.023 & 0.028 & 0.031 & 0.032 
	& 0.033& 0.034& 0.034& 0.035& 0.035 \\ \hline
$\langle \vec{k}^2 \rangle$ (GeV$^2$) 
			& 0.111 & 0.274 & 0.374 & 0.432 
	& 0.470 & 0.496 & 0.515 & 0.529 & 0.541 & 0.550 \\ \hline \hline
\end{tabular} 
\end{center} }
\vskip 0.5cm

Of course, as is well known, the HQET works well only at the scale
$\Lambda_{QCD} << \mu << m_b$. 
The results shown in Table I indicate that the implicit scale $\mu$ of our 
effective theory is not close to $m_b$.  On the other hand, 
for perturbation expansion to work, $\mu$ should not be too close  
$\Lambda_{QCD}$ either.  
Hence, we must choose a value of $\mu$ such that the corresponding
$\alpha_s \Lambda_{QCD}/m_Q$ is small enough to ensure the validity of 
the HQET. In Table I, we have also calculated $\langle \vec{k}^2 \rangle$, 
the mean square momentum carried by the quarks in the meson.
We find that $\langle \vec{k}^2 \rangle$ increases with increasing $n$,  
which indicates that in fact the scale $\mu$ 
also implicitly increases with $n$.
As a result, the corresponding $\alpha_s$ would be larger for smaller $n$. 
Thus, in order that higher order $1/m_Q$ corrections can be 
ignored,  we should choose a structure function $\Psi_n(v \cdot p_q)$ 
with a suitably larger $n$. 

In the following numerical analysis, we shall consider the scale with
\begin{equation} 
	\alpha_s (\mu) < 0.5~~ {\rm and}~ ~m_q (\mu) = 0.22 \sim 0.25 
		~~ {\rm GeV} ~~\, 
\end{equation}
and try to find a consistent set of parameters for each 
$\Psi_n (v\cdot p_q)$. 
The condition $\alpha_s (\mu) < 0.5$ guarantees that 
${\alpha_s (\mu) \Lambda_{QCD}/ m_b(\mu)}$ cannot be too large. 
Also, at this scale, we hope that the short-distance 
corrections given in Sec.~VII.A are still applicable.  
The range of the light quark mass we have assumed is close to the 
constituent quark mass often used in relativistic quark models. 
Now we proceed as follows.  For each $\Psi_n(v\cdot p_q)$, 
$\omega$ is fixed by $f_B=180$ GeV through Eq.~(\ref{fbdc}). 
We can then predict the HQET parameters $\overline\lambda_1$ and 
$\overline\lambda_2$.  Subsequently, $m_b(\mu)$,  
$\overline{\Lambda}(\mu)$, and $\alpha_s(\mu)$ are determined  
by fitting the experimental $\Delta M_{B^*B} = 0.046$ GeV \cite{PDG} 
and the experimental $B$-meson mass $M_B=5.279$ GeV \cite{PDG} 
via Eq.~(\ref{pbms}) and Eq.~(\ref{mass11}), respectively. 
We found that for $n \leq 3$, $\Psi_n(v\cdot p_q)$ allows no consistent 
fit to the experimental data. 
Only with $n > 4$, can we have good descriptions for 
various $B$ meson properties with small $1/m_b$ corrections.
The results are summarized in Table II.

\vskip 0.5cm
{\footnotesize
Table II. The parameter $\omega$ in the structure functions $\Psi_n
(v\cdot p_q)$ [(\ref{lwf})] and the bottom quark mass $m_b$ fitted to 
$f_B=180$ GeV and $\Delta M_{B^*B}=0.046$ GeV with a suitable giving 
scale by $\alpha_s(\mu)$ and $m_q(\mu)$ for each giving $n$ in 
Eq.~(\ref{lwf}), and all the predicted nonperturbative HQET parameters 
in the effective  theory, where $m_q, \omega, m_b$ and 
$\overline{\Lambda}$ are in unit
GeV, and $\langle \vec{k}^2 \rangle, \alpha_s \overline{\lambda}_1,
\lambda_1$ and $\lambda_2$ in unit GeV$^2$. 
\begin{center}
\begin{tabular}{|c|c|c|c|c|c|c|c|c|c|c|c|}  
\hline\hline $\Psi_n (v \cdot p_q)$ & ~$\alpha_s(\mu)$~ & ~$m_q(\mu)$~ &
~$\omega(\mu)~$ & ~$m_b(\mu)$~ &  ~$\langle \vec{k}^2
\rangle $ ~& ~$\alpha_s \overline{\lambda}_1$~ &~ $-{\lambda}_1$~ &  
${\lambda}_2$ ~& ~$\overline{\Lambda}$ & $m_c(\mu)$ & $g$ \\ \hline
$n=5$ & 0.457 & 0.251 & 0.405 & 4.907 & 0.292 & 0.254 & 0.038 & 0.087 & 
0.333 & 1.626 & 0.380 \\ \hline
$n=6$ & 0.431 & 0.248 & 0.671 & 4.901 & 0.329 & 0.265 & 0.064 & 0.089 & 
0.337 & 1.614 & 0.329 \\ \hline
$n=7$ & 0.417 & 0.245 & 0.937 & 4.894 & 0.352 & 0.271 & 0.081 & 0.090 & 
0.342 & 1.604 & 0.328  \\ \hline
$n=8$ & 0.408 & 0.242 & 1.205 & 4.888 & 0.368 & 0.275 & 0.093 & 0.090 & 
0.347 & 1.595 & 0.337 \\ \hline
$n=9$ & 0.401 & 0.238 & 1.477 & 4.880 & 0.381 & 0.278 & 0.103 & 0.091 & 
0.354 & 1.584 & 0.372 \\ \hline
$n=10$ & 0.395 & 0.234 & 1.752 & 4.869 & 0.391 & 0.280 & 0.111 & 0.091 & 
0.364 & 1.572 & 0.428 \\ \hline
$n=11$ & 0.390 & 0.230 & 2.029 & 4.859 & 0.399 & 0.282 & 0.117 & 0.091 & 
0.373 & 1.560 & 0.479 \\ \hline
$n=12$ & 0.385 & 0.226 & 2.310 & 4.844 & 0.406 & 0.283 & 0.123 & 0.091 & 
0.389 & 1.542 & 0.556 \\ \hline
exp. & --- & --- & --- & 4.89$^{**}$ & --- & --- & --- & 
0.090$^*$ & $ 0.33 \sim 0.45$\cite{Wise1} & 1.59$^{**}$ & $< 0.7$\cite{Exp1} 
\\ \hline\hline 
\end{tabular}
\end{center}
$^*$In the literature, the usual experimental value of $\lambda_2 \simeq 0.12$ 
GeV$^2$ is obtained by taking $m_b=M_B$ approximately. Here the 
experimental $\lambda_2\simeq 0.090$ GeV$^2$ is obtained by using $m_b^{\rm 
pole}=4.89$ GeV, the $b$-quark pole mass, in order to make a consistent 
comparison with the theoretical calculation used by Eq.~(\ref{msp}) with
the short-distance correction. 
\newline$^{**}$ There are no experimental data for the constituent quark 
masses $m_b$ and $m_c$. Here we list the bottom and charm quark pole masses
\cite{mass}} 
\vskip 0.5cm

From Table II, we see that $\langle \vec{k}^2\rangle$ increases with 
increasing $n$.  As noted earlier, this indicates that the scale $\mu$ 
in the phenomenological structure function 
$\Psi_n(v \cdot p_q)$ also goes up with $n$.  
Then to be consistent, the running strong coupling constant 
$\alpha_s (\mu)$  and 
the running quark masses $m_q(\mu)$ and $m_b(\mu)$ should decrease with
increasing $n$.  This expected behavior is clearly shown in 
Table II.  From table II, it is also interesting to see that the resulting 
$b$-quark mass is in between the constituent mass 
($\sim 4.8$ GeV) used in various 
model calculations and the pole mass ($4.89 \pm 0.05$ GeV). 
Furthermore, we also find that the hyperfine mass 
splitting  parameter $\lambda_2$ is quite stable and insensitive to the 
function $\Psi(v \cdot p_q)$.  This is not unexpected since $\lambda_2$ 
is directly related to the heavy meson mass splitting which is well 
measured.  The mass shift parameter $\lambda_1$ is very 
small in our calculation. In this work, $\lambda_1$ 
receives two contributions: one is the heavy quark kinetic energy 
($\langle \vec{k}^2 \rangle \simeq 0.3 \sim 0.4$ GeV$^2$ for
$n \geq 5 $) which is a negative contribution to 
$\lambda_1$ [see Eq.~(\ref{repara})], and the 
other is the chromo-electric interaction between the heavy quark 
and light degrees of freedom characterized by the parameter 
$\overline{\lambda}_1$ which is close to $0.3$ GeV$^2$ and is 
a positive contribution.  Therefore, the kinetic energy is 
largely compensated by the chromo-electric interaction energy, leading 
to a small and negative $\lambda_1$: $\lambda_1 = 0.04
\sim 0.11$ GeV$^2$ for $n=5 \sim 10$.  Then a fit to the physical masses 
of the $B$ mesons yield the nonperturbative HQET parameter 
$\overline{\Lambda}$: $\overline{\Lambda}\simeq
0.33 \sim 0.36$ GeV, which is consistent with results from
HQET analyses of the semi-inclusive $B$ decay data \cite{Wise1}. 
These numerical results provide a vote of confidence to the reliability 
of the effective theory proposed in this work.

For a further consistency check with HQET and HQS, we use 
the above results to calculate the $D$ meson observables. 
The $D$ meson mass is given by
\begin{equation}
	M_D = m_c (\mu) + \overline{\Lambda}(\mu) - {\lambda_1 \over 
		2 m_c(\mu)} - {3\over 4} \Delta M_{D^*D} \, ,
\end{equation}
where $\overline{\Lambda}$ and $\lambda_{1,2}$ are 
the same for both the $B$ and $D$ mesons. Then using 
the experimental $D$ mass data \cite{PDG}
\begin{equation}
	M_D= 1.8645 ~~ {\rm GeV}, ~~~ \Delta M_{D^*D} = 
		M_{D^*} - M_D = 0.142 ~~ {\rm GeV} \, , 
\end{equation}
we can uniquely determine the charmed quark mass. We get (see Table II)
\begin{equation}
	m_c(\mu) = 1.57 \sim 1.62 ~~ {\rm GeV}~(n=5\sim 10),
\end{equation}
which is close to the $m_c$ used in various quark model calculations, 
and also the pole mass $m^{\rm pole}_c=1.59 \pm 0.02$ GeV \cite{mass}.

Next we  proceed to calculate the Isgur-Wise function $\xi(v \cdot v')$ 
[cf. Eq.~(\ref{iwf1})], its slope parameter at the zero-recoil point
\begin{equation}
	\rho^2 = - \xi'(1) \, ,
\end{equation}
and the axial-vector coupling constant $g$ from Eq. (\ref{lfg}) or
Eq. (\ref{etg}). 
The Isgur-Wise function depends on the parameter $\omega$ and
the light quark mass $m_q$, and the axial-vector coupling constant 
also depends on $\overline{\Lambda}$. Hence, once  
$\omega$ and $\overline{\Lambda}$ are determined for a given $m_q$, 
$\xi(v\cdot v')$, $\rho^2$, and $g$ can all be predicted. In Fig.~4, we 
plot the Isgur-Wise 
function as a function of $v \cdot v'$.  It is remarkable to 
see that the structure functions $\Psi_n (v \cdot p_q)$ (\ref{lwf}) 
with different $n$ give very similar results. 
In particular, the Isgur-Wise function obtained from $\Psi_{n=10}$
is almost identical to that form the light-front wave 
functions (\ref{gaus}) and (\ref{wfus}). The slope parameter of the 
Isgur-Wise function at the zero-recoil point
is found to be: \begin{equation}
	\rho^2 = 1.1 \sim 1.2 ~~~ (n=5 \sim 10)\, , 
\end{equation}
which is also consistent with other theoretical analyses \cite{Neubert94}.

Having determined $\overline{\Lambda}$ from the heavy meson mass 
calculations, we can now predict the axial-vector coupling constant $g$. 
As listed in Table II, we have 
\begin{equation}
	g=  0.33 \sim 0.43 ~~~(n = 5 \sim 10) \, , 
\end{equation}
which is consistent with the experimental constraint $g < 0.7$ 
\cite{cheng92,Exp1} 
derived from $D^* \rightarrow D + \pi$ decays.  Our result is also 
close to the QCD sum rule results, which tend to concentrate within the 
range $0.2<g<0.4$.

\section{Conclusions and perspective}

In this paper, we have given a detail account of a recently proposed  
field theoretical description of heavy mesons \cite{cheung97a}. 
The effective theory incorporates 
heavy quark symmetry and the heavy quark effective theory, from
which a natural realization of heavy mesons in the heavy quark 
limit as a composite particle of the reduced heavy quark coupled 
with a brown muck of light degrees of freedom is provided. 
This theory is fully covariant, so that Feynman diagrammatic 
techniques can be use to carry out perturbative calculations.
Moreover the effective theory preserves the simplicity of a 
conventional quark model, in fact, for those quantities which do not have 
the so-called $Z-$diagram contributions, light-front quark model results 
can be reproduced as a special case in the present theory. 
Thus this theory provides a link between the fundamental QCD and the 
phenomenologically successful quark model, so that the difficult 
subject of hadronic bound state physics can be quantitatively studied in 
a covariant framework. 

The effective theory provides a quasi-first-principles description of 
the heavy meson dynamics.  
Although at present the description of the 
heavy mesons structure function $\Psi(v \cdot p_q)$  
is still phenomenological in nature, 
it offers a systematic approach 
to evaluate the $1/m_Q$ corrections based on the first-principles 
$1/m_Q$ expansion of QCD. 
This resembles very much the situation of the QCD analysis of deep 
inelastic scatterings, in which the low
energy dynamics (described by parton distribution functions)
is determined phenomenologically and perturbative corrections
are given in a fully first-principles way.  Here the situation may
even be better since the phenomenological part is constrained by 
HQS and  HQET, and the nonperturbative QCD dynamics should
be much simpler in the heavy quark limit.

As we have seen, the introduction of the structure function
$\Psi(v\cdot p_q)$ is crucial for
a field theoretic realization of the heavy mesons as composite
particles. In fact, this structure function is related to the 
covariant wave functions of heavy meson bound states, and it
essentially describes the brown muck structure of the light
degrees of freedom inside the heavy mesons in the heavy quark
limit. 
Our results show that it is not possible to extend the popular  
Gaussian-type wave functions to be used 
in a fully covariant formulation.  
However, we find that a Lorentzian-type function 
can be readily adopted in a covariant formulation for heavy 
meson structures, and numerically it gives a very good description 
of the physical properties of heavy mesons. The predictions 
for all the HQET parameters are consistent  with experiment. 
$\overline{\Lambda}$ is about $0.33 \sim 0.39$ GeV [for $\Psi_n( v\cdot 
p_q)$ with $n=5 \sim 12$] which is consistent with results from HQET 
analyses of the experimental data\cite{Wise1}.  
The heavy quark masses $m_b$ and $m_c$, determined 
by fitting to the $B^*-B$ and $D^*-D$ mass differences, are 
about $4.84 \sim 4.91$ GeV and $1.54\sim 1.63$ GeV respectively, which 
agree with those used in various relativistic quark 
model calculations, and also their respective pole
masses.  We find that the mass shift 
parameter $\lambda_1$ in HQET is negative and very small  
(about -0.04 $\sim -0.12$) because of 
a large cancellation between the heavy quark kinetic energy and the 
chromo-electric interaction between the heavy quark and the light degrees 
of freedom. The Isgur-Wise function we obtained is also consistent 
with other calculations. Thus, we have self-consistently determined 
and predicted all the HQET parameters within the effective theory.

We expect that the effective theory presented in this paper 
can now be applied to describe various heavy meson processes,
such as the inclusive $B$ decays and various
exclusive $B$ decays without involving light mesons. 
For processes involving light mesons, we must first extend
our framework so that light mesons are also included.  
Finally, to solve the structure function $\Psi(v \cdot p_q)$ 
from the fundamental theory in the heavy quark limit is an important 
and interesting challenge.  
We shall leave these topics to future investigations.

\acknowledgements

This work is supported in part by the National Science Council of
the Republic of China under Grants NSC87-2112-M001-048,
NSC87-2112-M001-002 and NSC86-2816-M001-008L.



\newpage
\begin{center} {\bf Figure Captions}  \end{center}
\begin{description}
\item[Fig.~1]~~ {Heavy-light quark scattering in  (a) quark-quark 
  coupling picture, and (b) meson-quark coupling picture, which 
  determine the composite particle structure of heavy mesons.}
\item[Fig.~2]~~ {Feynman diagrams for (a) the Isgur-Wise function, (b)
  decay constant, and (c) strong axial-coupling constant of
  heavy mesons.}
\item[Fig.~3]~~ {Feynman diagrams for $1/m_Q$ corrections to 
  heavy meson masses.}
\item[Fig.~4]~~ {The Isgur-Wise functions as a function of $v \cdot v'$
  that obtained from the wave functions $\Psi_n(v \cdot p_q)$, ($n=6, 8, 
  10)$ [(\ref{lwf})] and compare with the Isgur-Wise functions obtained 
  from the light-front wave functions $\Psi^G$ [(\ref{gaus})] and $\Psi^M$ 
  [(\ref{wfus})] \cite{cheng97b}.} 
\end{description}

\newpage
\begin{figure}
    \centerline{\epsffile{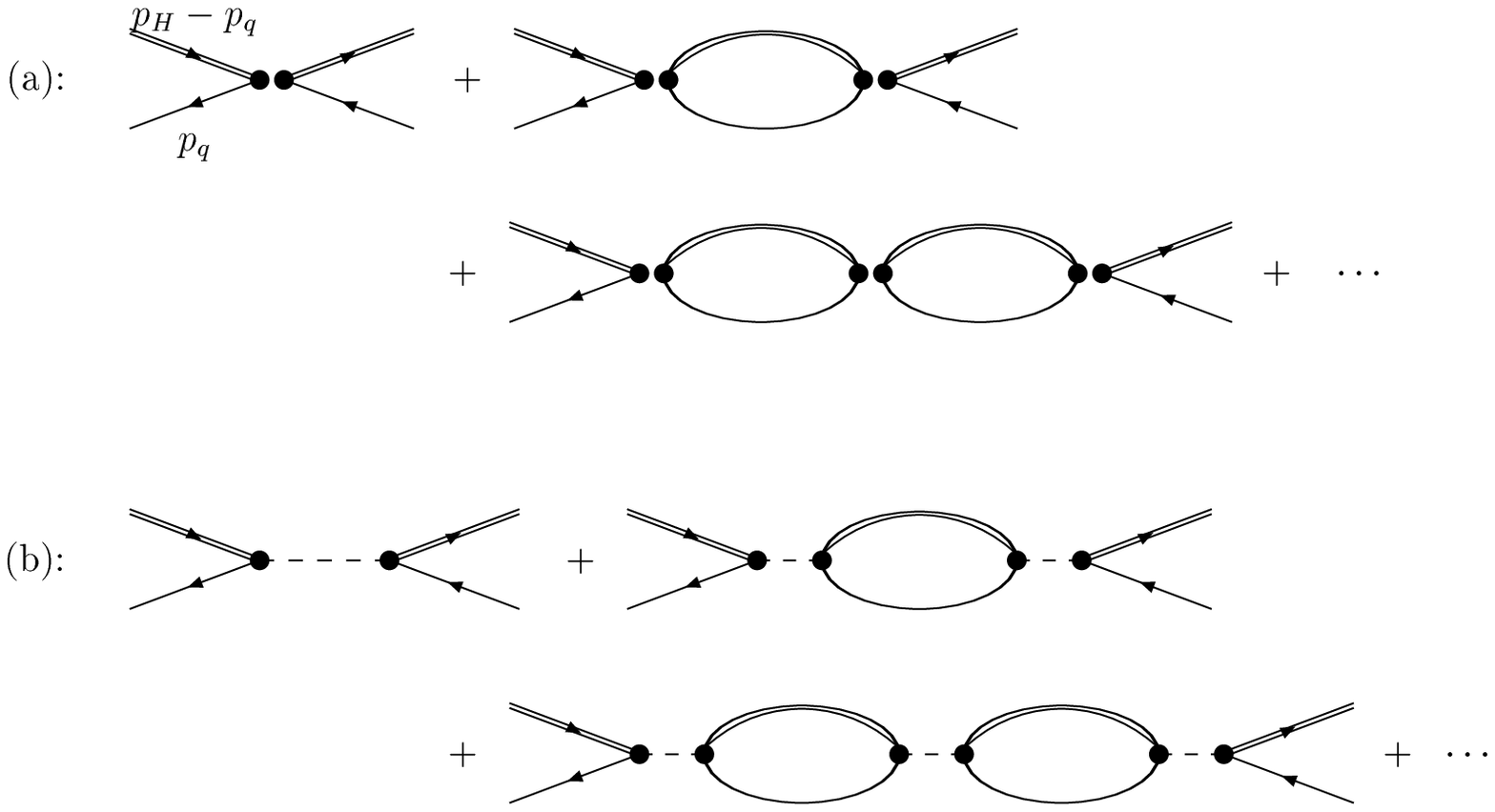}}
\vspace{-12cm}
   FIG.~1. {\small Heavy-light quark scattering in  (a) quark-quark 
     coupling
     picture, and (b) meson-quark coupling picture, which determine the
     composite particle structure of heavy mesons.}
    \label{fig1} 
\end{figure}

\newpage
\begin{figure}
    \centerline{\epsffile{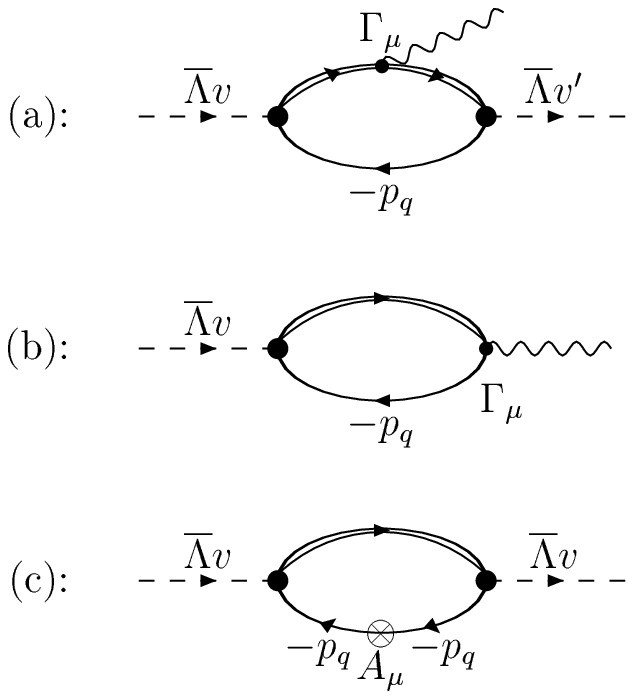}}
\vspace{-15cm}
    FIG.~2. {\small Feynman diagrams for (a) the Isgur-Wise function, (b)
    decay constant, and (c) strong axial-coupling constant of
    heavy mesons.}
\vspace{1cm}
    \label{fig2} 
\end{figure}

\newpage
\begin{figure}
    \centerline{\epsffile{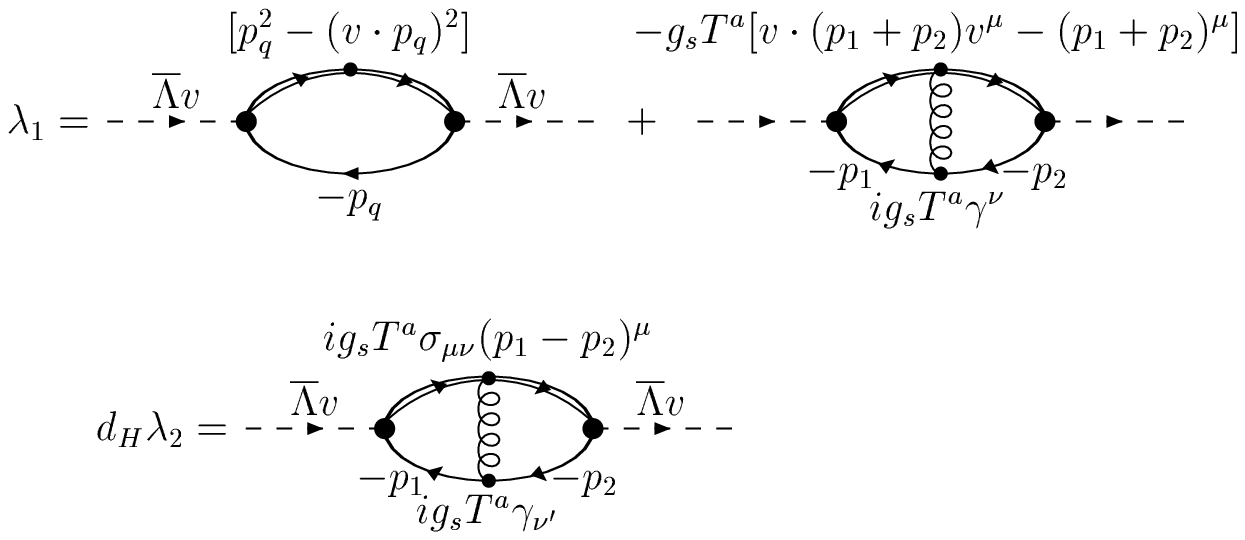}}
\vspace{-15cm}
    FIG.~3. {\small Feynman diagrams for $1/m_Q$ corrections to 
     heavy meson masses.}
    \label{fig3} 
\end{figure}

\newpage
\begin{figure}[ht]
\epsfxsize=13cm
    \centerline{\epsffile{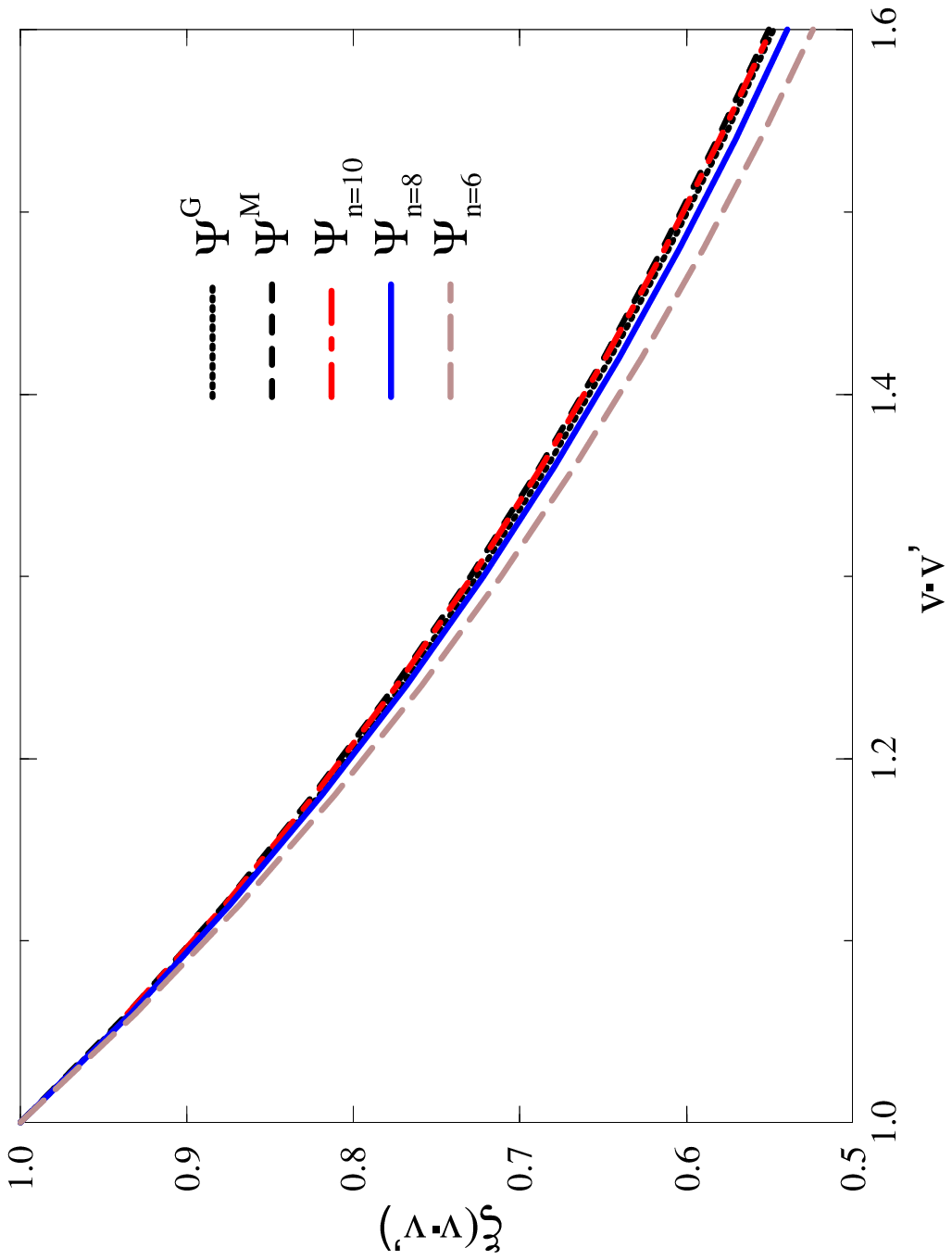}}
\vspace{-0cm}
 FIG.~4. {\small The Isgur-Wise functions as a function of $v \cdot v'$
that obtained from the wave functions $\Psi_n(v \cdot p_q)$, ($n=6, 8, 
10)$ [(\ref{lwf})]
and compare with the Isgur-Wise functions obtained from the light-front
wave functions $\Psi^G$ [(\ref{gaus})] and $\Psi^M$ [(\ref{wfus})] 
\cite{cheng97b}. }   \label{fig4}

\end{figure}


\begin{references}
    \bibitem{IW89} N. Isgur and M. B. Wise, Phys. Lett. {\bf B232},
                113 (1989); {\bf B237}, 527 (1990).
    \bibitem{Georgi90} H. Georgi, Phys. Lett. {\bf B240}, 447 (1990);
    		E. Eichten and B. Hill, Phys. Lett. {\bf B234}, 511
                (1990); {\bf B243}, 427 (1990).
    \bibitem{Neubert94} M. Neubert, Phys. Rep. {\bf 245}, 261 (1994);
    		Also see hep-ph/9801269.
    \bibitem{lattice} C. Bernard, Y. Shen, and A. Soni, Phys. Lett. {\bf
		B317}, 164 (1993); A. Abada et al. (ELC Collaboration),
		Nucl. Phys. {\bf B416}, 675 (1994); C. R. Allton et
		al. (APE Collaboration),
                Phys. Lett. {\bf B345}, 513 (1995); K. C. Bowler et al.
                (UKQCD Collaboration),
		Phys. Rev. {\bf D52}, 5067 (1995); Nucl. Phys. {\bf B461}, 
		327 (1996).
    \bibitem{zhang97} W. M. Zhang, Phys. Rev. {\bf D56}, 1528 (1997).
    \bibitem{Isgur89}N. Isgur, D. Scora, B. Grinstein, and M. B. Wise,
		Phys. Rev. {\bf D 39}, 799 (1989); D. Scora and N. Isgur, 
		Phys. Rev. {\bf D52}, 2783 (1995); H. Y. Cheng and B. Tseng,
		Phys. Rev. {\bf D53}, 1457 (1996). 
    \bibitem{MIT} M. Sadzikowski and K. Zalewski, Z. Phys. {\bf C59},
		677 (1993).
    \bibitem{Ball91} P. Ball, V. M. Braun, and H. Dosch, Phys. Rev. {\bf D44},
		3567 (1991); P. Ball, Phys. Rev. {\bf D48}, 3190 (1993);
		V. M. Belyaev, A. Khodjamirian, and R. Ruckl, Z. Phys.
		{\bf C60}, 349 (1993); T. Huang and C.-W. Luo, Phys. Rev. 
		{\bf D50}, 5775 (1994); A. Ali, V. M. Braum, and H. Simma, 
		Z. Phys. {\bf C83}, 437 (1994); P. Colangelo, F. De Fazio,
		P. Santorelli, and E. Scrimieri, Phys. Rev. {\bf D53},
		3672 (1996).
    \bibitem{Te76} W. Jaus, Phys. Rev. {\bf D 41}, 
		3394 (1990); {\bf 44}, 2851 (1991); {\bf 53}, 1349 (1996);
    		P. J. O'Donnell and Q. P. Xu, Phys. Lett. {\bf B325},
		219 (1994); {\bf 336}, 113 (1994); P. J. O'Donnell, Q. P. 
		Xu, and H. K. K. Tung, Phys. Rev. {\bf D52}, 3966 (1995). \\
		For a review on light-front dynamics, see   W. M. Zhang, 
		Chin. J. Phys. {\bf 32}, 717 (1994).
   \bibitem{dubin93} A. Dulib and A. Kaidalov, Yad. Fiz. {\bf 56}, 164 
		(1993), [Phys. At. Nucl. {\bf 56}, 237 (1993)].
   \bibitem{cheung97} C. Y. Cheung, C. W. Hwang, and W. M. Zhang, Z. Phys.
		{\bf C75}, 657 (1997).  
   \bibitem{cheng97a} H. Y. Cheng, C. Y. Cheung, and C. W. Hwang, Phys.
		Rev. {\bf D55}, 1559 (1997).
   \bibitem{demchuk97} N. B. Demchuk, P. Yu. Kulikov, I. M. Narodetskii
		and P. J. O'Donnell, Phys. At. Nucl. {\bf 60}, 1292 (1997).
   \bibitem{cheng97b} H. Y. Cheng, C. Y. Cheung, C. W. Hwang, and W. M.
		Zhang, hep-ph/9709412, to appear in Phys. Rev. {\bf D57},
		(1998). 
   \bibitem{cheung97a} C. Y. Cheung and W. M. Zhang, submitted to Phys. 
		Lett. , hep-ph/9712258.
    \bibitem{BSW} M. Wirbel, S. Stech, and M. Bauer, Z. Phys. {\bf C29},
    \bibitem{HNZ} R. Hagg, Phys. Rev. {\bf 112}, 669 (1958); K. Nishijima,
	{\it ibid}, {\bf 111}, 995 (1958); {\bf 122}, 298 (1961); R.
	L. Zimmermann, {\it ibid}, {\bf 141}, 1124 (1964).  
    \bibitem{Lurie} D. Lurie, {\it Particles and Fields} (Interscience,
     	New York, 1969).
    \bibitem{Yan94} T. M. Yan, H. Y. Cheng, C. Y. Cheung, G. L. Lin, Y. C. 
		Lin, and H. L. Yu, Phys. Rev. {\bf D 46}, 1148 (1992); 
		M. B. Wise, Phys. Rev. {\bf D 45}, R2188 (1992); 
		G. Burdman and J. Donoghue, Phys. Lett. {\bf B 280}, 
		287 (1992).
    \bibitem{cheng92} H. Y. Cheng, C. Y. Cheung, G. L. Lin, Y. C. Lin, T. M.
		Yan, and H. L. Yu, Phys. Rev. {\bf D 46}, 5060 (1992); 
		{\it ibid}. {\bf D 47}, 1030 (1993); 
    \bibitem{Falk} A. Falk {\it et al.,} Nucl. Phys. {\bf B343}, 1 (1990).
    \bibitem{BSUV} I.I. Bigi, M.A. Shifman, N.G. Uraltsev, 
                   and A.I. Vainshtein, Phys. Rev. {\bf D50} 2234 (1994).
    \bibitem{PDG} Particle Data Group, Phys. Rev. {\bf D54}, 1 (1996).
    \bibitem{Neubert96} M. Neubert, Int. J. Mod. Phys. {\bf A11}, 4173 (1996).
    \bibitem{Exp1} ACCMOR Collaboration, S. Barlag {\it et al.} Phys.
		Lett. {\bf B278}, 480 (1992).
    \bibitem{Wise1} M. Gremm and I. Stewart, Phys. Rev. {\bf D55}, 1226 
		(1997); A. F. Falk, M. Luke and M. J. Savage, Phys. Rev,
		{\bf D53}, 6316 (1996).
    \bibitem{mass} H. Fusaoka and Y. Koide, hep-ph/9712201.
\end{references}
\end{document}